\documentclass[fleqn,usenatbib]{mnras}

\usepackage[T1]{fontenc}

\DeclareRobustCommand{\VAN}[3]{#2}
\let\VANthebibliography\thebibliography
\def\thebibliography{\DeclareRobustCommand{\VAN}[3]{##3}\VANthebibliography}

\usepackage{amsmath,amsfonts,amssymb,amscd,wasysym,graphicx,cleveref}
\usepackage{pdfpages}
\usepackage{changepage}

\graphicspath{{./}{figures/}}

\newcommand\Gemini{\textit{Gemini}}
\newcommand\gnirs{GNIRS}

\newcommand{\spitz}{\textit{Spitzer}}
\newcommand{\wise}{\textit{WISE}}
\newcommand{\sdss}{\textrm{SDSS}}

\newcommand\progID{GN-2017B-Q-51}

\newcommand{\pypeit}{\texttt{PypeIt}}

\newcommand{\oi}{[\textrm{O}~\textsc{i}]}
\newcommand{\oii}{[\textrm{O}~\textsc{ii}]}
\newcommand{\oiii}{[\textrm{O}~\textsc{iii}]}

\newcommand{\civ}{[\textrm{C}~\textsc{iv}]}
\newcommand{\feii}{\textrm{Fe}~\textsc{ii}}

\newcommand{\ha}{${\rm H\alpha}$}
\newcommand{\hb}{${\rm H\beta}$}

\newcommand{\mum}{$\mu$m}

\newcommand{\Mbh}{$M_{\textrm{BH}}$}
\newcommand{\LIR}{$L_{5\micron}$}
\newcommand{\Lbol}{$L_{\textrm{bol}}$}
\newcommand{\zbest}{$z_{\textrm{best}}$}
\newcommand{\HaHb}{${\rm H\alpha}/{\rm H\beta}$}

\title[\Gemini/\gnirs\ $z\sim2$ obscurerd quasars]{Infrared spectroscopic confirmation of $z\sim2$ photometrically-selected obscured quasars}

\newcommand{\jhu}{The William C. Miller III Department of Physics and Astronomy, Johns Hopkins University, Baltimore, MD 21218, USA}

\author[Y. Ishikawa et al.]{Yuzo Ishikawa,$^{1}$\thanks{E-mail: yishika2@jhu.edu}
Ben Wang,$^{2,3}$
Nadia L. Zakamska,$^{1,4}$
Gordon T. Richards,$^{5}$
Joseph F. Hennawi,$^{3,6}$
\newauthor
Angelica B. Rivera,$^{7}$
\\
$^{1}$\jhu\\
$^{2}$Department of Astronomy, Tsinghua University, Beijing 100084, China\\
$^{3}$Leiden Observatory, Leiden University, P.O. Box 9513, 2300 RA Leiden, The Netherlands\\
$^{4}$Institute for Advanced Study, Princeton University, Princeton, NJ 08544, USA\\
$^{5}$Department of Physics, Drexel University, 32 S.\ 32nd Street, Philadelphia, PA 19104, USA\\
$^{6}$Department of Physics, Broida Hall, UC Santa Barbara, Santa Barbara, CA 93106-9530, USA\\
$^{7}$Department of Physics, US Naval Academy, 572C Holloway Rd., Annapolis, MD 21402, USA\\
}

\date{Accepted XXX. Received YYY; in original form ZZZ}

\pubyear{2022}

\begin{document}
\label{firstpage}
\pagerange{\pageref{firstpage}--\pageref{lastpage}}
\maketitle

\begin{abstract}
The census of obscured quasar populations is incomplete, and remains a major unsolved problem, especially at higher redshifts, where we expect a greater density of galaxy formation and quasar activity. We present Gemini GNIRS near-infrared spectroscopy of 24 luminous obscured quasar candidates from the Sloan Digital Sky Survey's Stripe 82 region. The targets were photometrically selected using a \wise/W4 selection technique that is optimized to identify IR-bright and heavily-reddened/optically-obscured targets at $z>1$. We detect emission lines of \ha, \hb, and/or \oiii\ in 23 sources allowing us to measure spectroscopic redshifts in the range $1<z<3$ with bolometric luminosities spanning $L=10^{46.3}-10^{47.3}$ erg s$^{-1}$. We observe broad $10^3-10^4$ km s$^{-1}$ Balmer emissions with large \HaHb\ ratios, and we directly observe a heavily reddened rest-frame optical continuum in several sources, suggesting high extinction ($A_V\sim7-20$ mag). Our observations demonstrate that such optical/infrared photometric selection successfully recovers high-redshift obscured quasars. The successful identification of previously undetected red, obscured high-redshift quasar candidates suggests that there are more obscured quasars yet to be discovered.  
\end{abstract}

\begin{keywords}
galaxies: high-redshift - quasars: general - quasars: supermassive black holes - quasars: emission lines - infrared:galaxies
\end{keywords}



\section{Introduction} \label{sec:intro}
According to the standard unification model, many of the observed properties of active galactic nuclei (AGN) may be explained by differences in the viewing angle with respect to the dusty torus \citep[e.g.,][]{Antonucci1993ARAA}. In this scheme, there is a compact source of optical, ultraviolet (UV) and X-ray emitting continuum associated with the accretion disk and the corona near the supermassive black hole. On larger scales, other components include the broad line region, the dusty obscuring torus, and the narrow line region of gas photo-ionized by the quasar radiation \citep[e.g.,][]{KrolikBegelman1988,LawrenceElvis2010}. In this picture, Type-1 quasars are observed with a direct line of sight to the central source revealing the broad emission lines. Type-2 quasars are observed sideways through the obscuring torus, so that only narrow (and forbidden) emission lines and infrared (IR) emission produced by the reprocessing in the torus are observed. 

In addition to the purely orientation-based hypothesis for the difference between obscured and unobscured sources, there are other possible interpretations. Popular galaxy evolution models postulate that unobscured and obscured quasars represent different phases of evolution. In this picture, the obscured phase precipitates a phase of strong mechanical feedback, during which the obscuring material is expelled to reveal the unobscured quasar \citep[e.g][]{Sanders1988ApJ, CanalizoStockton2001, Hopkins2006, Glikman2012}.  There is evidence that some obscured sources, although a minority, may be caused by transient absorbers in the line of sight \citep[e.g.,][]{LaMassa2015}. Some of the most luminous obscured quasars at high redshifts like extremely red quasars (ERQs)  may be super-Eddington accretors \citep[e.g.,][]{Assef2015, Perrotta2019MNRAS, Zakamska2019}. These quasars may be obscured by inflowing and outflowing materials. Quantifying and characterizing obscured quasars remains an ongoing problem in modern astrophysics.

A complete understanding of the growth of supermassive black holes requires a complete census of both the unobscured and obscured quasar populations. The unification models predict that the number density of unobscured and obscured quasars should be comparable through cosmic time, peaking at $z\sim2.5$ \citep{Richards2006, Ross2013}. A large population of obscured quasars would affect our existing calculations of the accretion efficiency of supermassive black holes \citep{Soltan1982MNRAS,YuTremaine2002}. 
Furthermore, obscured quasars are a key component to the cosmic X-ray background \citep[e.g.,][]{Treister2005,Lanzuisi2009}. Moreover, $z\sim2$ marks the epoch of rapid growth of the supermassive black holes through mass accretion in the most luminous quasars, so it is critical to characterize the quasars at high redshift.

Decades of work have established that the number density of unobscured and obscured quasars at low redshifts ($z<1$) are comparable \citep{LawrenceElvis2010,reyes2008AJ}. 
However, the census at high redshift remains incomplete and debated \citep[e.g.,][]{Polletta2008, Eisenhardt2012ApJ, Lusso2013ApJ777, Hennawi2013ApJ766, Prochaska2013ApJ776, Assef2015}.  
Some X-ray studies suggest that the obscured fraction may be a function of luminosity with an anti-correlation between obscured fraction and quasar luminosity, yet remain debated \citep[e.g.,][]{ueda2003, Glikman2018}. 
The apparent paucity may imply that obscured quasars are not very common, possibly due to physical changes in their obscuring medium 
\citep{Hasinger2008AA, Lawrence1991MNRAS, KoniglKartje1994, Elitzur2006ApJ}. Alternatively, they may have eluded discovery due to incomplete selection. 

It is important to highlight the difficulty of observing obscured quasars. Quasars typically outshine their host galaxies by several orders of magnitude in the rest-frame UV and optical wavelengths. However, a key characteristic of obscured quasars is that dust obscuration in the line of sight can heavily redden the quasar continuum. As a result, the largest catalogs of obscured quasars are largely limited to low redshift targets that are optically selected based on their extremely strong narrow emission lines which originate outside of the obscuring material \citep[e.g.,][]{zakamska2003AJ,reyes2008AJ,Yuan2016MNRAS}. Alternatively, powerful X-ray detections can be indicative of quasar activity. However, the X-ray selection is typically limited to sources with modest column densities that allow hard X-rays to travel unimpeded. At extremely high, Compton-thick column densities, the absorption makes X-ray detections prohibitively difficult, despite high intrinsic luminosities of the quasar hidden behind the obscuration \citep[e.g.,][]{Goulding2018ApJ,Zappacosta2018, LaMassa2019ApJ,Ishikawa2021ERQ}. Therefore, a more robust strategy is to take advantage of the dust obscuration by probing the mid-infrared (MIR) emission, which is thought to be both more universal and isotropic, compared to UV/optical selections. Even Compton-thick quasars should be MIR-bright due to the reprocessed emission from the warm/hot dust in/around the obscuring torus. However, caution is needed as this method has many contaminants. For example, MIR continuum emission can also result from intense star-forming galaxies; the clumpy nature of the torus, which can allow MIR emission to escape unimpeded; and dust in the polar regions rather than the obscuring torus \citep{HickoxAlexander2018}. 

MIR-based color selection using \wise\ \citep{Wright2010} or \spitz\ \citep{Werner2004} has had successes in identifying obscured quasars. Recent work \citep{Eisenhardt2012ApJ, Wu2012ApJ,Yan2013AJ145, Ross2015MNRAS453, Glikman2018} demonstrated that a combined selection of bright mid-IR sources that are optically faint efficiently identifies obscured quasars both at $z<1$ and at $z\sim2$.  However, canonical mid-IR \spitz/IRAC color selection wedges, which rely on $8\micron$ (rest-frame $2.7\micron$) fluxes, do not sufficiently probe the peak of the quasar-heated dust emission at $\sim10\micron$ \citep{Lacy2004ApJS,Stern2005ApJ, Polletta2008,Donley2012ApJ,Lacy2013ApJS} and may miss populations of  high redshift obscured quasars. 
Therefore, it would be ideal to probe even longer wavelengths, such as with the all-sky \wise/W4 ($\sim22\micron$) coverage, complemented with \spitz\ observations. In this paper, we take advantage of color-selection strategies based on both \wise/W4 and \spitz\ fluxes to select luminous $z\sim2$ obscured candidates over an area larger than those of the previous searches.

This paper is structured as follows. In Section \ref{sec:dataRedux} we discuss the sample selection and data reduction. In Section \ref{sec:specAnaly} we present the analysis of spectra and photometry of our candidates. And Section \ref{sec:discuss} we discuss the implications of our results, and we conclude in Section \ref{sec:concl}. We use the AB magnitude system unless otherwise specified. We adopt the $h = 0.7$, $\Omega_M = 0.3$, and $\Omega_{\Lambda} = 0.7$ cosmology. 

\section{Target selection}\label{sec:dataRedux}
\subsection{Sample selection}\label{sec:sample}

\begin{figure}
      \includegraphics[width=0.97\columnwidth]{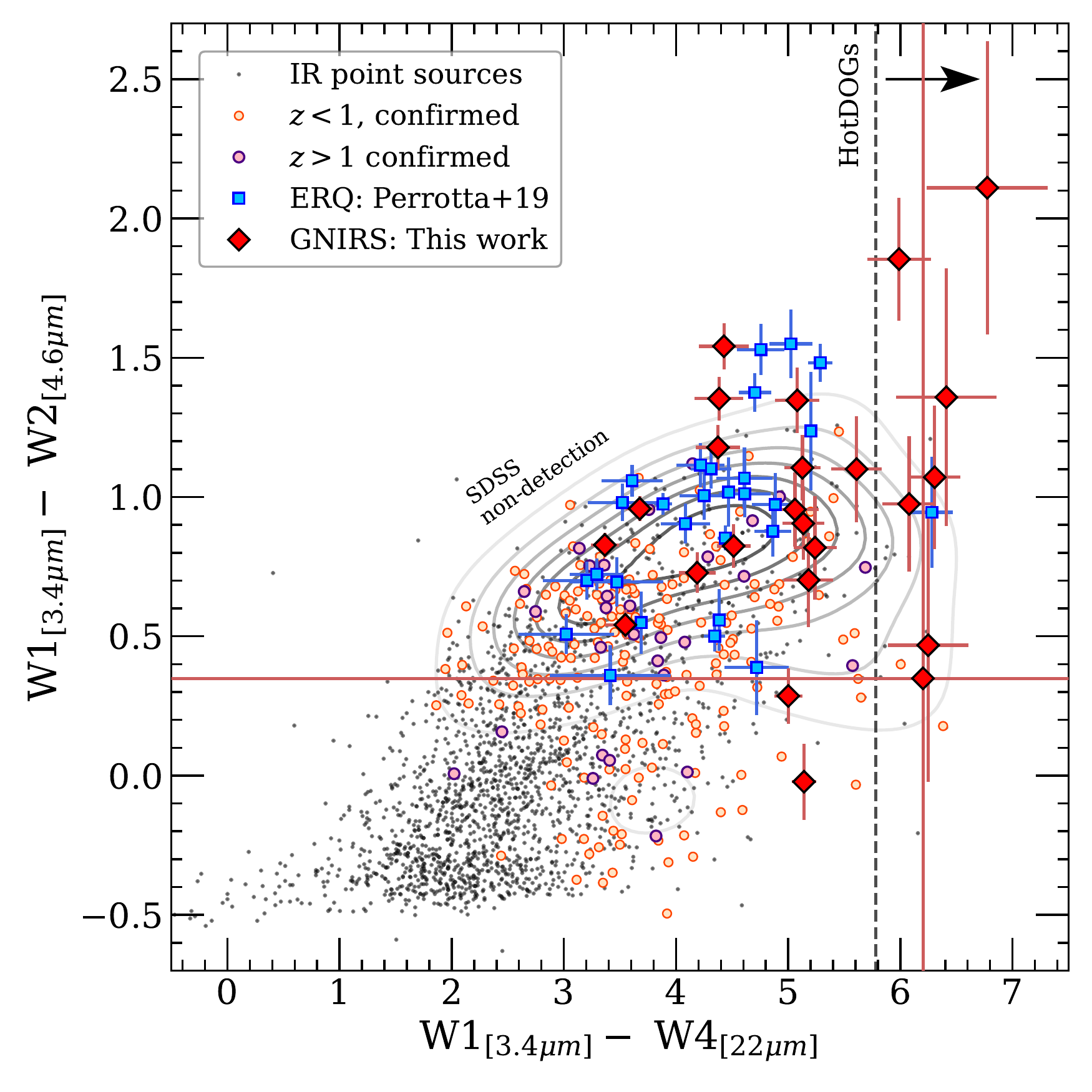} 
	 \caption{The color-color distribution of the \wise/W4 selected objects. The black points show the distribution of $f_{24\micron}>1\textrm{ mJy}$ point sources detected in \spitz/IRAC. The orange and purple circles indicate spectroscopically confirmed $z<1$ and $z>1$ sources, respectively. The gray contours indicate the \sdss\ non-detections. The red diamonds indicate our sources. We stress that our selections in this study primarily rely on \wise/W4 and SDSS fluxes. Any suggestive color-color relations may serve as a guide for future searches of obscured quasars. We also compare with the color cuts of HotDOGs (\citealt{Eisenhardt2012ApJ}; black dashed line) and the observed \wise\ colors of ERQs (\citealt{Perrotta2019MNRAS}; blue squares): both of which rely on color-based selections.}
	 \label{fig:colors} 
\end{figure}


Our selection uses \wise, \spitz, and SDSS data to identify sources that are MIR luminous, yet optically faint. First, \wise/W4 bright sources are selected. We require $5\sigma$ detection with $6 < W4 < 8\textrm{ Vega mag}$ ($f_{W4} \geq 5\textrm{ mJy}$). The bright limit is applied to reduce stellar contamination. Additional criteria are applied to ensure robust MIR detections and to limit contamination from stars and other spurious sources: $5\sigma$ detection of \wise/W3 ($\sim12\micron$) and $W1 \geq 12\textrm{ Vega mag}$. We also consider the \wise\ data-quality flags to avoid artifacts ($\texttt{w4flg}=0$ and $\texttt{w3flg}=0$) and to limit to point-sources ($\texttt{ext\_flg}=0$ and $\texttt{nb}\leq2$). These \wise-selected sources are then cross-matched with \spitz\ surveys to obtain precise astrometry of $\sim0.1"$ for narrow-slit spectroscopic follow-up. 

The MIR selections are further narrowed down to identify optically-faint sources. 
The full search area includes fields from SHELA \citep{Papovich2016ApJS224}, SWIRE \citep{Lonsdale2003}, SERVS \citep{Mauduit2012}, COSMOS \citep{Sanders2007}, SDWFS, and AllWISE surveys. For this \Gemini/\gnirs\ study, we limit the selection to optically faint sources from the \spitz\ IRAC Equatorial Survey (SpIES; PI Richards), which is a SWIRE-depth survey in the \sdss\ Stripe 82 area \citep{Timlin2016ApJS}. Stripe 82 is the largest field considered. 
The effective search area is $\gtrsim 164 \textrm{ deg}^2$. We define optical faintness as either having $r>23\textrm{ AB mag}$ or undetected in the Sloan Digital Sky Survey (\sdss; \citealt{york2000}). If we combine the SDSS and \wise+\spitz\ selections, we obtain a $r-W4 >13.5$ color cut. The effective color cut implied by these criteria is somewhat more stringent than that for the selection of Dust Obscured Galaxies (DOGS; \citealt{Dey2008ApJ}). Previous studies \citep{Brand2007, Ross2015MNRAS453} show that this selection uncovers luminous obscured quasars at $z\sim2$. 

We identified 373 sources from the W4 selection over all fields considered. This selection recovers 15 known $z<1$ sources and 22 known $z>1$ sources, of which 18 are known $z>1$ Type-2 or reddened Type-1 quasars, including a $z=3.106$ HotDOG \citep{Lacy2013ApJS, Glikman2013, Banerji2015, Tsai2015, Glikman2018}. From the candidates, we selected sources that lack spectroscopic confirmation. We find 26 optically faint sources in the Stripe 82 region that are appropriate for \Gemini/\gnirs. Interestingly, only 3 sources are detected in \sdss\ photometry. We obtained \Gemini/\gnirs\ spectra for 24 of these sources to identify the \ha\ emission line to obtain spectroscopic redshifts. 

In Figure \ref{fig:colors} we show a color-color plot of the sources to compare with other known color-selected obscured quasars. We stress that our selections do not rely on $W1-W2$ or $W1-W4$ colors. We discuss the implications of the selection criteria in Section \ref{sec:color}. We list our \Gemini/\gnirs-observed targets and their selection properties in Table \ref{tab:targetlist}. We summarize the selection criteria here:
\begin{equation}
    \begin{split}
    & W4 \geq  5 \textrm{ mJy} \\
    & r > 23 \textrm{, or undetected in SDSS}.
    \end{split}
\end{equation}

\begin{table*}
	\centering
	\caption{Sample list of observed quasars. The target coordinates are determined from \spitz+\wise\ and SDSS photometry. The targets are ordered by coordinates. Sources observed on different dates have been co-added to produce a single spectrum. We list the selection properties. We classify the sources as A/B/BB/C/D based on their continuum and emission line strengths. All continuum detections have a red slope; `BB' marks sources with a blue excess. $^{\dagger}$ denotes sources with no SDSS detections; SDSS r-band upper limit is $r>23 \textrm{ AB mag}$ \citep{york2000}. }
	\label{tab:targetlist}
	\begin{tabular}{lcclccc}
	    \hline
        Target name & RA       & DEC     &  Date(s) observed & W4 [22\micron]   &  $r'_{\textrm{SDSS}}$   & Category  \\ 
                    & (J2000)  & (J2000) &  (yyyy-mm-dd)     & (AB mag)         & (AB mag)                &    \\ 
        \hline
    J002407.02$-$001237.2	&	00:24:07.02	    &	$-$00:12:37.2   &	2017-10-06          	&  $ 7.39 \pm 0.12 $   &    -$^{\dagger}$     &      C    \\
    J004157.77$-$002932.1	&	00:41:57.77	    &	$-$00:29:32.1	&	2017-10-07            	&  $ 7.94 \pm 0.21 $   &    -$^{\dagger}$     &      BB   \\
    J004729.25$+$000358.8	&	00:47:29.25	    &	$+$00:03:58.8	&	2017-10-21, 2017-10-30	&  $ 7.67 \pm 0.16 $   &    -$^{\dagger}$     &      D    \\
    J005424.45$+$004750.2	&	00:54:24.45	    &	$+$00:47:50.2	&	2017-10-21            	&  $ 7.21 \pm 0.13 $   &    -$^{\dagger}$     &      C    \\
    J010552.86$-$002351.2	&	01:05:52.86	    &	$-$00:23:51.2	&	2017-10-21, 2017-10-30	&  $ 7.78 \pm 0.16 $   &    -$^{\dagger}$     &      C    \\
    J011222.64$-$001633.0	&	01:12:22.64	    &	$-$00:16:33.0	&	2017-11-05            	&  $ 7.73 \pm 0.20 $   &    -$^{\dagger}$     &      C    \\
    J011314.49$+$002917.1	&	01:13:14.49	    &	$+$00:29:17.1	&	2017-11-05            	&  $ 7.85 \pm 0.19 $   &    -$^{\dagger}$     &      C    \\
    J013033.47$+$000950.4	&	01:30:33.47	    &	$+$00:09:50.4	&	2017-11-05            	&  $ 7.98 \pm 0.20 $   &    -$^{\dagger}$     &      C    \\
    J014939.96$+$005256.7	&	01:49:39.96	    &	$+$00:52:56.7	&	2017-11-07            	&  $ 7.95 \pm 0.17 $   &    -$^{\dagger}$     &      B    \\
    J015055.28$+$005600.2	&	01:50:55.28	    &	$+$00:56:00.2   &	2017-11-06          	&  $ 7.98 \pm 0.21 $   &    -$^{\dagger}$     &      D    \\
    J015235.29$-$002459.4	&	01:52:35.29	    &	$-$00:24:59.4	&	2017-11-06            	&  $ 7.26 \pm 0.09 $   &    $22.57\pm0.18$    &      A    \\
    J021345.44$+$002436.1	&	02:13:45.44	    &	$+$00:24:36.1	&	2017-11-07            	&  $ 7.86 \pm 0.16 $   &    -$^{\dagger}$     &      A    \\
    J021426.98$-$000021.3	&	02:14:26.98	    &	$-$00:00:21.3	&	2017-11-07, 2017-11-09	&  $ 7.56 \pm 0.12 $   &    -$^{\dagger}$     &      C    \\
    J021514.76$+$004223.8	&	02:15:14.76	    &	$+$00:42:23.8   &	2017-09-27          	&  $ 7.39 \pm 0.12 $   &    -$^{\dagger}$     &      B   \\
    J022127.60$+$005024.6	&	02:21:27.60	    &	$+$00:50:24.6   &	2017-09-29          	&  $ 7.70 \pm 0.14 $   &    -$^{\dagger}$     &      C    \\
    J222920.83$+$002253.5	&	22:29:20.83	    &	$+$00:22:53.5	&	2017-11-07            	&  $ 7.04 \pm 0.11 $   &    -$^{\dagger}$     &      C    \\
    J223358.38$-$000414.9	&	22:33:58.38	    &	$-$00:04:14.9	&	2017-11-05            	&  $ 7.81 \pm 0.19 $   &    -$^{\dagger}$     &      D    \\
    J223904.01$-$003054.9	&	22:39:04.01 	&	$-$00:30:54.9	&	2018-01-02           	&  $ 7.46 \pm 0.16 $   &    -$^{\dagger}$     &      D    \\
    J223911.98$-$005422.3	&	22:39:11.98	    &	$-$00:54:22.3	&	2017-09-29, 2017-09-30	&  $ 7.75 \pm 0.18 $   &    $23.31\pm0.32$    &      D    \\
    J224338.04$+$001749.9	&	22:43:38.04    	&	$+$00:17:49.9	&	2017-09-30            	&  $ 7.51 \pm 0.17 $   &    -$^{\dagger}$     &      A    \\
    J225851.90$-$002207.0	&	22:58:51.90 	&	$-$00:22:07.0   &	2017-10-01             	&  $ 7.89 \pm 0.20 $   &    -$^{\dagger}$     &      C    \\
    J225956.84$-$000918.4	&	22:59:56.84 	&	$-$00:09:18.4	&	2018-01-03              &  $ 7.76 \pm 0.19 $   &    -$^{\dagger}$     &      B    \\
    J232925.01$+$002057.7	&	23:29:25.01 	&	$+$00:20:57.7	&	2017-11-07           	&  $ 7.88 \pm 0.19 $   &    -$^{\dagger}$     &      C    \\
    J233441.49$+$003114.0	&	23:34:41.49 	&	$+$00:31:14.0	&	2017-11-06           	&  $ 7.62 \pm 0.16 $   &    $23.10\pm0.22$    &      BB    \\
        \hline
    \end{tabular}
\end{table*}

\subsection{Observation and reduction}\label{sec:obsredux}
\label{sec:obsredux}

The targets were observed with the \Gemini/\gnirs\ spectrograph \citep{Elias2006SPIE6269E14E, Elias2006SPIE6269E4CE} under \progID\ (PI: Richards) over several nights spanning 2017-09-27 to 2018-01-03. We used GNIRS with the $0.15"/\textrm{pixel}$ camera and the $32 \textrm{ l/mm}$ grating in the cross dispersed mode for a coverage over 0.9-2.4 \micron. We used the $0.45"$ slit for an effective spectral resolution of $R\sim1100$. Observations were performed with the ABBA observing sequence. 
Each object has a total integration time of 2400s. 

We reduce the data with the \pypeit\ pipeline \citep{pypeit:joss_arXiv, pypeit:joss_pub, pypeit:zenodo}. This reduction produces fully calibrated science spectra. For objects with clear continuum, we run the automatic spectra extraction routine. For objects with faint or no continuum detected, we manually look for emission lines to select the extraction aperture in the 2D spectra for the spectral extraction. Flux calibration is applied with standard stars observed on the same night. Telluric corrections are also applied. 

\begin{figure*}
\begin{tabular}{l}
\begin{tabular}{c}
\includegraphics[width=0.95\textwidth]{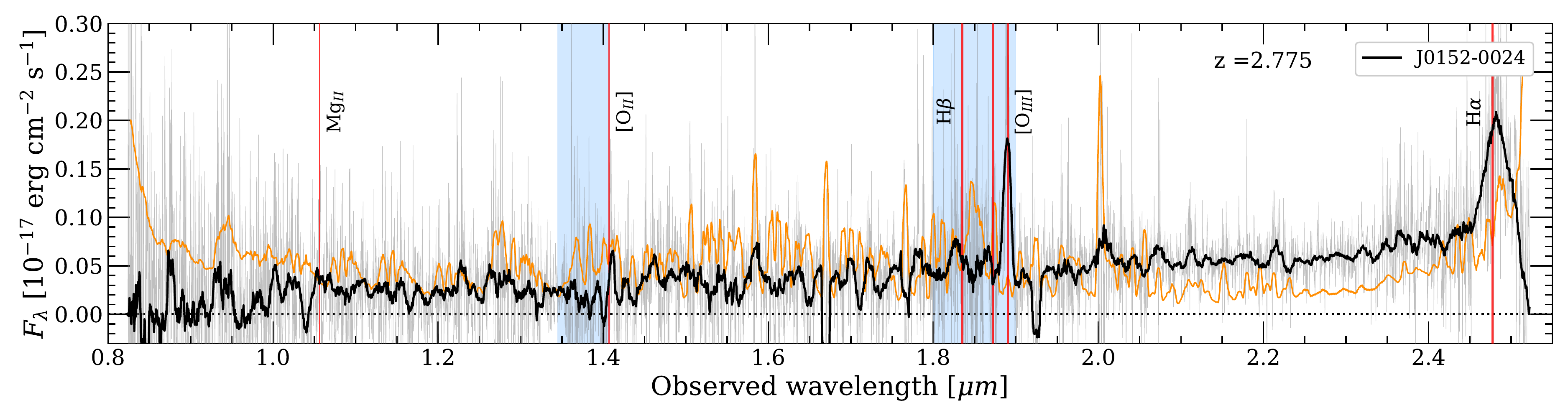} \\
\includegraphics[width=0.95\textwidth]{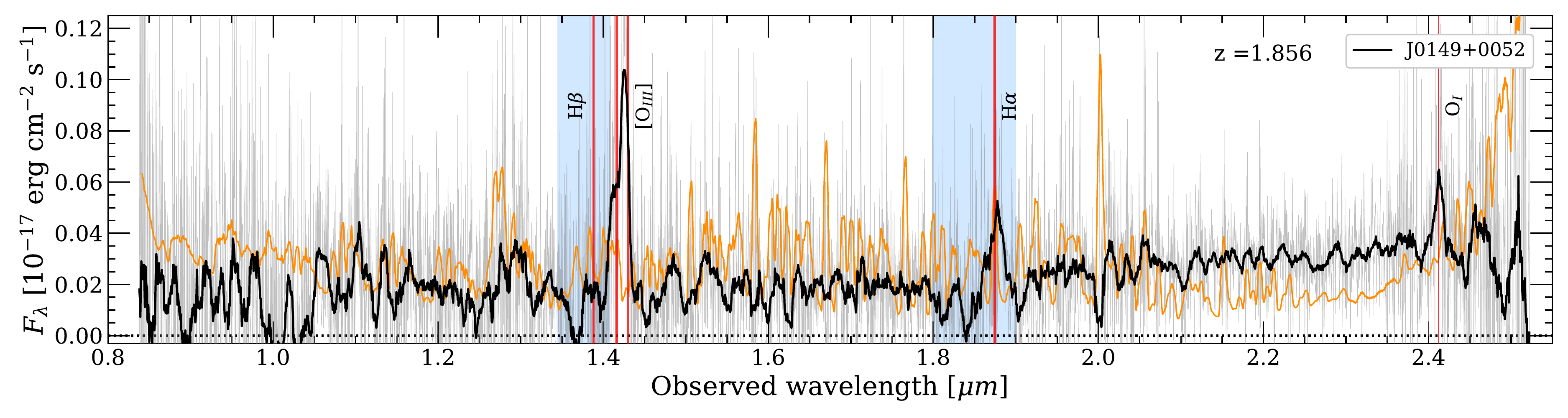} \\
\includegraphics[width=0.95\textwidth]{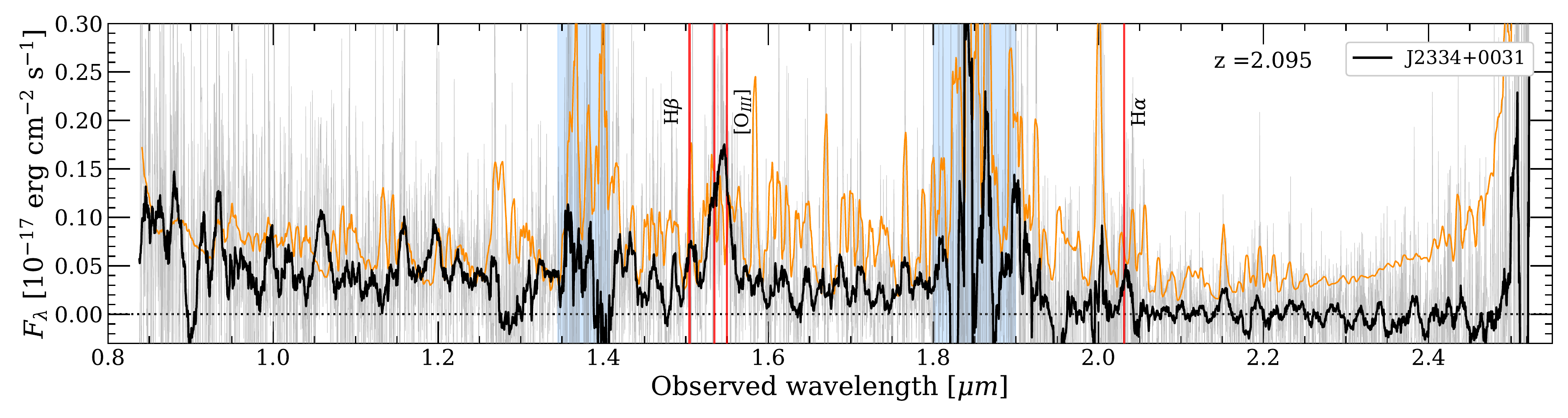} \\
\includegraphics[width=0.95\textwidth]{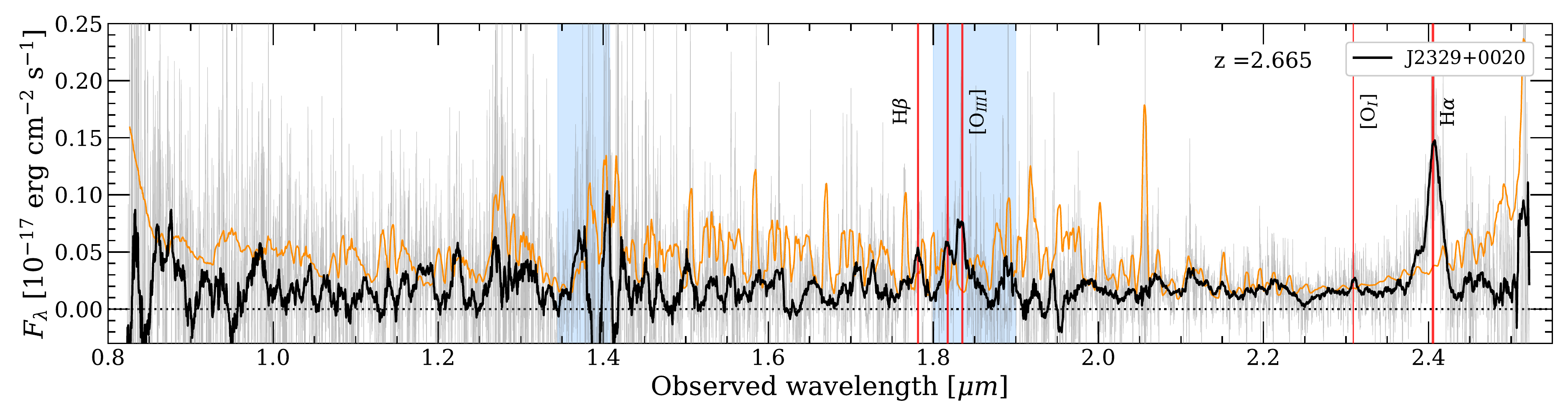} \\ 
\end{tabular}
\end{tabular}
\caption{Example spectra of A/B/BB/C sources with continuum and/or emission lines detected. We do not show the featureless D-sources here. The shaded blue regions mark telluric absorption, and the orange curve is the noise vector plot. Raw spectra are in light gray and the smoothed spectra in bold black. The vertical red lines indicate the emission lines identified.}
\label{fig:spect_show} 
\end{figure*}

We examine the fully reduced and calibrated spectra to categorize the sources as A/B/C/D based on the observed continuum and emission line strengths. The sources are classified into the following categories: (A) strong line(s) with strong continuum, (B) weak line(s) with weak continuum, (BB) `B'-type with a blue-sloped excess, (C) weak line with no continuum, and (D) no line and no continuum. 
We indicate these classifications in Table \ref{tab:targetlist}. We are able to confidently categorize most sources, although some required additional Keck/LRIS observations for confirmation (Wang et al., in prep). Roughly half of the observed sources have significant emission lines and/or continuum. The other half show emission with low equivalent widths, which made it difficult to determine the spectroscopic redshift. Only one source (J0047+0003) showed no features. We show a subset of the observed sources with significant emission and/or continuum in Figure \ref{fig:spect_show}.

\section{Results: spectral analysis}\label{sec:specAnaly}
\begin{table*}
\begin{adjustwidth}{-1cm}{}
	\centering
	\caption{Emission line properties of the targets listed by coordinates. $^*$ indicates sources with significant telluric absorption of \ha. $^{\dagger}$ indicates sources with featureless spectra, but are spectroscopically confirmed by Keck/LRIS observations; \zbest\ is the best redshift estimate, and $z_{\textrm{GNIRS}}$ is the redshift estimate using only Gemini/GNIRS data. $^{\textrm{w}}$ indicate redshifts identified determined by Gemini/GNIRS and Keck/LRIS (Wang et al., in prep). }
	\label{tab:spectprop}
	\begin{tabular}{llccccccccc}   
	    \hline
        Target & \zbest\  & $z_{\textrm{G}}$  &  Ref.line &
$\log_{10}L_{H\alpha}$ & FWHM$_{H\alpha}$  &
$\log_{10}L_{H\beta}$  & FWHM$_{H\beta}$  &
$\log_{10}L_{[\textrm{O}~\textsc{iii}]}$  & FWHM$_{[\textrm{O}~\textsc{iii}]}$ 
& $\log_{10}L_{\textrm{bol}}$  \\ 
       &   &   &  &
($\textrm{erg s}^{-1}$)  & ($\textrm{km s}^{-1}$)  & 
($\textrm{erg s}^{-1}$)  & ($\textrm{km s}^{-1}$)  & 
($\textrm{erg s}^{-1}$)  & ($\textrm{km s}^{-1}$)  & 
($\textrm{erg s}^{-1}$)  \\
        \hline
J0024$-$0012	    &  1.528$^{\textrm{w}}$& 0.925  & \oiii\ & $42.01\pm 0.06$  &  $2355\pm 193$  &  $41.5\pm 2.0$   &  $650\pm 300$ & $42.33\pm 0.15$  &  $3300\pm 900$ &  $46.97\pm 0.02$ \\
J0041$-$0029    	&  2.09$^{\textrm{w}}$ & 1.495  & \ha\ & $42.61\pm 0.01$  &  $2780\pm 1700$  &  -  &  - &  - &  - &  $46.71\pm 0.05$ \\
J0047$+$0003        &  -  & - & - & -      &  -  &  -  &  - &  - &  - &  - \\
J0054$+$0047    	&  2.17 & 2.17  & \ha\ & -      &  -  &  -  &  - &  - &  - &  $47.31\pm 0.03$ \\
J0105$-$0023$^{\dagger}$  &  1.865 & 1.865  & \ha\  &  $44.33\pm 0.05$     &  $200\pm 70$  &  -  &  - &  - &  - &  $46.81\pm 0.05$ \\
J0112$-$0016        &  2.99$^{\textrm{w}}$ & -  & \oiii\  & -      &  -  &  -  &  - &  $41.88\pm 0.07$ &  $2000\pm 1000$ &  $47.72\pm 0.03$ \\
J0113$+$0029    	&  2.33 & 2.33 & \ha\ & $42.38\pm 0.21$  &  $470\pm 175$  &  -  &  - &  - &   &  $46.83\pm 0.10$ \\
J0130$+$0009    	&  2.5   & 2.5 & \ha\ & $42.70\pm 0.13$  &  $1180\pm 255$  &  -  &  - &  - &  - &  $47.23\pm 0.04$ \\
J0149$+$0052    	&  1.85 & 1.85 & \ha\ & $42.06\pm 0.09$  &  $2370\pm 450$  &  -  &  - &  $42.43\pm 0.05$ &   $1840\pm 100$  &  $46.95\pm 0.04$ \\
J0150$+$0056     	&  1.7    & 1.7 & \ha\ & $42.76\pm 0.39$  &  $3300\pm 1000$  &  -  &  - &  - &  - &  $46.91\pm 0.04$ \\
J0152$-$0024	    &  2.775  & 2.775& \ha\ & $43.37\pm 0.01$  &  $3840\pm 300$  &  -  &  - &  $42.90\pm 0.18$ &   $940\pm 100$  &  $47.35\pm 0.04$ \\
J0213$+$0024$^*$    &  1.805  & 1.805& \oiii\ & $43.05\pm 0.01$  &  $6896\pm 300$  &  $42.25\pm 0.05$  &  $5000\pm 1000$ &   $42.38\pm 0.20$  & $2386\pm 260$ &  $47.02\pm 0.03$ \\
J0214$-$0000	    &  1.627  & 1.627& \ha\ & -      &  -  &  -  &  - &  - &  - &  $47.03\pm 0.02$ \\
J0215$+$0042	    &  0.88   & 0.88& \ha\ & $42.38\pm 0.03$  &  $2150\pm 100$  &  $41.47\pm 0.39$  &  $1690\pm 1400$ &  $42.13\pm 0.12$ &   $3000\pm 200$  &  $46.27\pm 0.01$ \\
J0221$+$0050	    &  2.48   & 2.48& \ha\ & $42.94\pm 0.02$  &  $4400\pm 370$  &  -  &  - &  - &  - &  $46.25\pm 0.04$ \\
J2229$+$0022	    &  1.93   & 1.93& \ha\ & $42.58\pm 0.12$  &  $2840\pm 1190$  &  -  &  - &  $42.19\pm 0.13$ &   $950\pm 197$  &  $47.19\pm 0.03$ \\
J2233$-$0004$^{\dagger}$  &  1.602$^{\textrm{w}}$  & -& \ha\ & -      &  -  &  -  &  - &  - &  - &  $46.23\pm 0.16$ \\
J2239$-$0030$^{\dagger}$  &  1.905$^{\textrm{w}}$  & -& \ha\ & -      &  -  &  -  &  - &  - &  - &  $47.20\pm 0.03$ \\
J2239$-$0054$^{\dagger}$  &  2.085$^{\textrm{w}}$  & -& \ha\ & -      &  -  &  -  &  - &  - &  - &  $47.12\pm 0.04$ \\
J2243$+$0017$^*$	&  1.905  & 1.905& \oiii\ & -  &  -  &  -  &  - &  $44.7\pm 0.05$ &   $1865\pm 100$  &  $47.10\pm 0.04$ \\
J2258$-$0022        &  2.42   & 2.42& \ha\ & $42.29\pm 0.05$  &  $2210\pm 400$  &  -  &  - &  - &  - &  $46.95\pm 0.08$ \\
J2259$-$0009    	&  1.885  & 1.885& \ha\ & $42.09\pm 0.09$  &  $2350\pm 115$  &  -  &  - &  $42.24\pm 0.12$ &  $1750\pm 100$  &  $47.03\pm 0.04$ \\
J2329$+$0020	    &  2.665  & 2.665& \ha\ & $43.18\pm 0.03$  &  $2340\pm 115$  &  $42.24\pm 0.15$  &  $677\pm 150$ &  $42.62\pm 0.11$ &  $1330\pm 280$ &  $47.16\pm 0.06$ \\
J2334$+$0031    	&  2.095$^{\textrm{w}}$  & 1.355 & \ha\ & $42.19\pm 0.09$  &  $4350\pm 200$  &   $42.17\pm 0.28$  &  $1600\pm 000$ &  $42.85\pm 0.02$ &  $3500\pm 500$ &  $46.95\pm 0.06$ \\
        \hline
    \end{tabular}
 \end{adjustwidth}
\end{table*} 

\subsection{Redshift estimates}\label{sec:zshift}

\begin{figure}
	 \begin{center}
	 \begin{tabular}{c}
    \includegraphics[width=0.95\columnwidth]{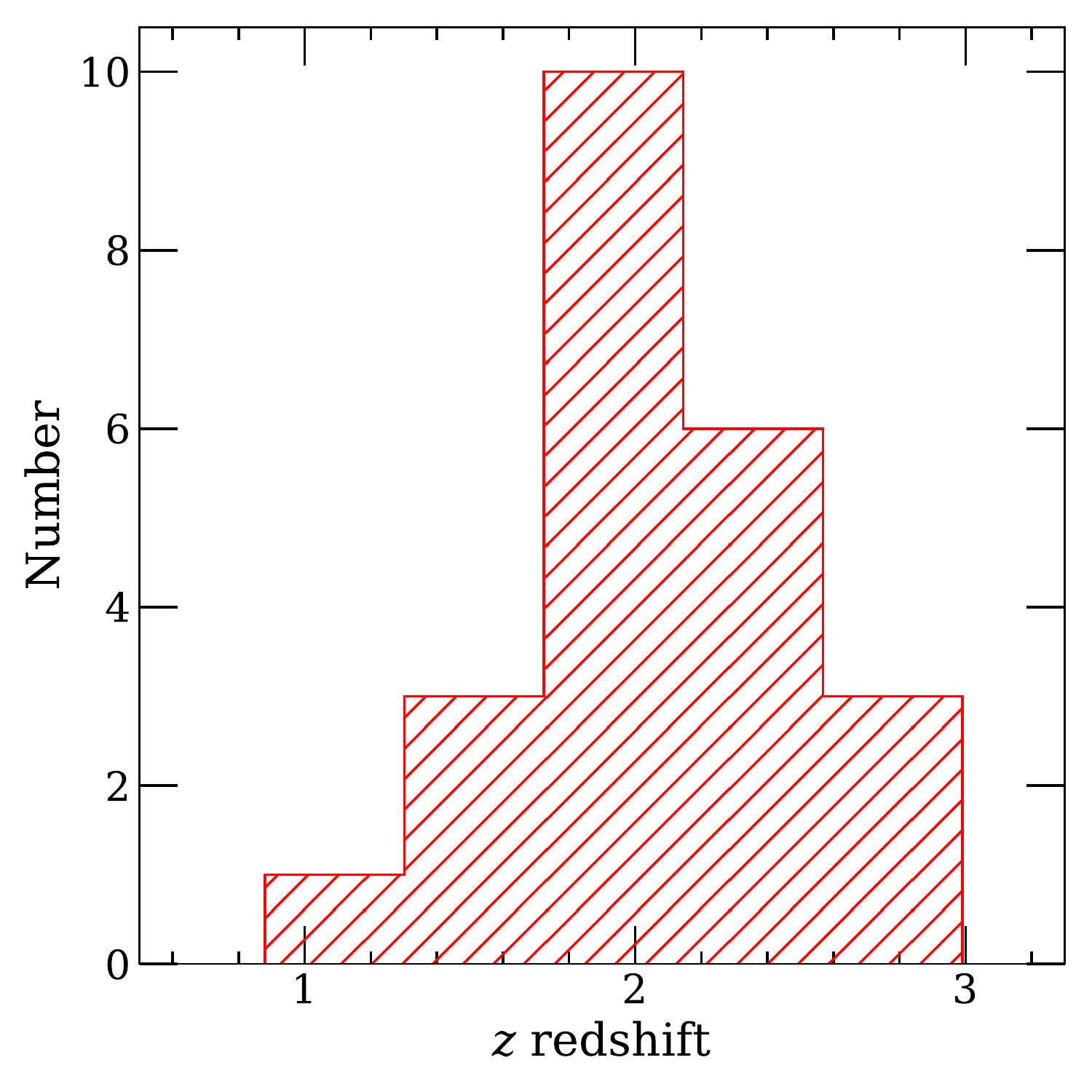}
	 \end{tabular}
	 \end{center}
	 \caption{Redshift distribution of the observed sources. } 
	 \label{fig:zhist} 
\end{figure}

We compute the redshifts in two steps. First, we visually identify emission lines of interest (e.g. \ha\ and \hb$+$\oiii) to determine the rough spectroscopic redshift. We set these as priors to fit Gaussians to the observed emission lines. We then refine the spectroscopic redshift with the best-fit line centroid shifts. To this end, we first subtract out the underlying continuum. All sources with strongly detected continuum flux have a red slope. We define the continuum by masking out the emission lines and telluric contamination at $\sim1.4\micron$ and $\sim1.9\micron$ and interpolating over the masked spectrum. Finally, we fit single Gaussians to the continuum-subtracted emission line spectrum. 

For targets with sufficient \gnirs\ detections, we take the \ha\ and/or \hb$+$\oiii\ lines to estimate the spectroscopic redshifts. The observed equivalent widths of the emission lines vary. For sources with heavy telluric contamination, we use other emission lines for spectroscopic redshift determination. For example, J0213+0024 and J2243+0017 show heavy $\sim1.9\micron$ telluric contamination, making it difficult to accurately find the centroid of the \ha\ line; instead we use the \oiii\ doublet and \oi $\lambda8446$\AA. If we only detect one emission line (e.g. J2334+0031), then we assume it is \ha. We fit the \oiii\ doublet by kinematically tying the lines to \oiii5009\AA\ and setting the peak flux \oiii\ line ratio $\lambda5009$\AA/$\lambda4959$\AA=2.993. Some sources were originally identified as featureless  `D'-sources, but were confirmed with Keck/LRIS observations (Wang et al., in prep). For example, the \ha\ line for J0112-0016 is redshifted out of the GNIRS coverage, and the one line visible in the GNIRS spectrum turns out to be \oiii.  

In Table \ref{tab:spectprop}, we show the redshift estimates from \Gemini/\gnirs-only data, $z_{\textrm{GNIRS}}$, and the ``best'' estimates from the combined \Gemini/\gnirs+Keck/LRIS data, \zbest. The differences in the redshift estimates highlight how challenging it can be to identify obscured quasars. In Figure \ref{fig:zhist} we show the redshift distribution of the spectroscopically confirmed objects. We can see that the initial color selection preferentially selects $1<z<3$ objects with the average redshift of $\langle z_{\textrm{best}}\rangle= 2.008\pm0.45$. We do not see any redshift dependence on the spectral types outlined in Section \ref{sec:obsredux}.

\subsection{Emission line properties}\label{sec:emline}

\begin{figure*}
	 \begin{center}
	 \begin{tabular}{c}
     \includegraphics[width=0.98\textwidth]{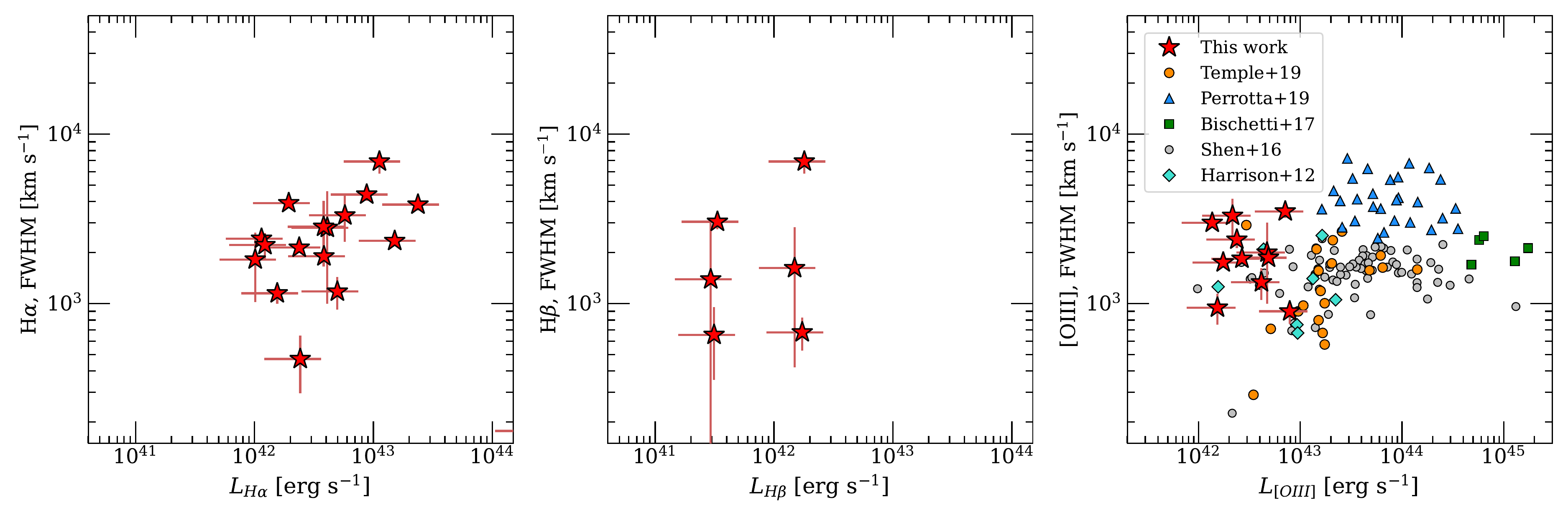}
	 \end{tabular}
	 \end{center}
	 \caption{Line kinematics and the line luminosity of \ha\ (left), \hb\ (center), and \oiii\ (right). The red symbols indicate sources with sufficient signal-to-noise ratio (SNR) for line fits. The observed \ha\ display broad widths similar to typical quasars. However, the width distribution of \oiii\ and \ha\ are inconsistent at $\sim1\sigma$ using the KS-test. We also compare the \oiii\ properties with values from literature \citet{Harrison2012,ZakamskaGreene2014, Shen2016,Bischetti2017, Perrotta2019MNRAS,Temple2019MNRAS}. No obvious correlation between the line luminosity and widths is seen. } 
	 \label{fig:lineCompare} 
\end{figure*}

\begin{figure*}
	 \begin{center}
	 \begin{tabular}{c}
    \includegraphics[width=0.98\textwidth]{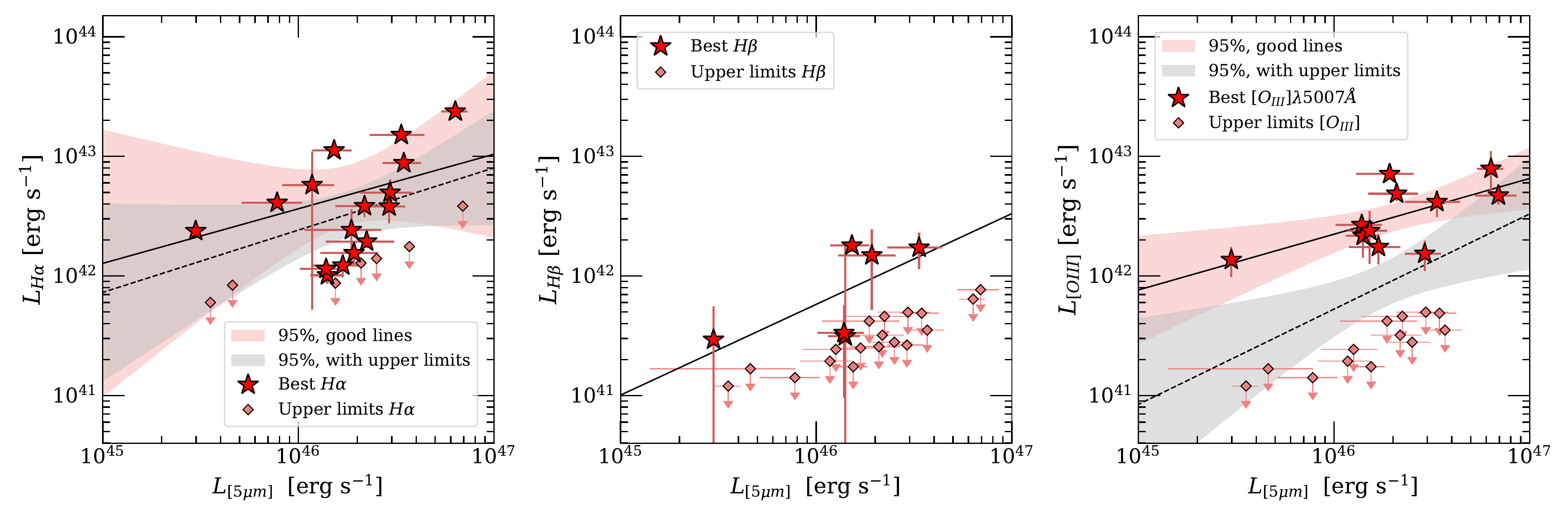} \\
    \includegraphics[width=0.98\textwidth]{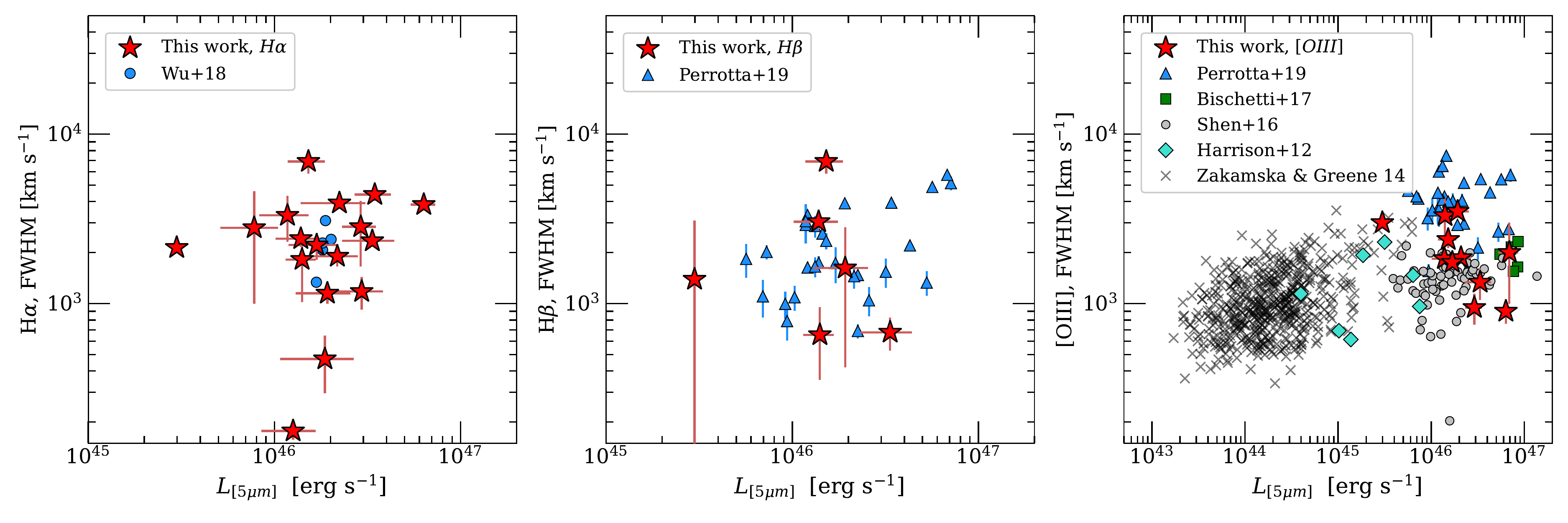}
	 \end{tabular}
	 \end{center}
	 \caption{Line luminosities (top row) and kinematics (bottom row) shown against \LIR\ for \ha\ (left), \hb\ (center), and \oiii\ (right). The red symbols indicate sources with sufficient SNR for line fits. Light-red diamonds indicate the $1\sigma$ upper limits based on the average continuum flux.  We also compare the \ha, \hb\, and \oiii\ kinematics with values from \citet{Harrison2012, ZakamskaGreene2014, Shen2016,Bischetti2017, Wu2018,Perrotta2019MNRAS}. For the comparison with our sample, we converted the published $w_{80}$ velocities to FWHM. Our sources show a wide range of velocities with 3 reaching ERQ-like velocities (see bottom right).}
	 \label{fig:lineVwise} 
\end{figure*}

After we obtain the best-fit line parameters, we calculate the line luminosities. The line luminosity is calculated as the observed fluxes multiplied by $4\pi D_L^2$ assuming that the emission is isotropic, where $D_L$ is the luminosity distance at \zbest. In Table \ref{tab:spectprop} we show the calculated line luminosity and the full-width-half-maximum (FWHM; $2.355\sigma$). The \ha\ line is the brightest with luminosity reaching $10^{42}\textrm{-}10^{43.4} \textrm{ erg s}^{-1}$, whereas the \hb$+$\oiii\ lines are nearly $\times10$ fainter. The observed \ha\ fluxes are commensurate with those of DOGS \citep{Brand2007}. This is also at the limit of GNIRS sensitivities, based on GNIRS ITC predictions of emission from a $z=2.3$ source with $1\times10^{-16}\textrm{ erg s}^{-1}\textrm{ cm}^{-2}$ and $\textrm{FWHM}\times1000\textrm{ km s}^{-1}$ superposed on a continuum of $1\times10^{-18}\textrm{ erg s}^{-1}\textrm{ cm}^{-2}$\AA$^{-1}$.

In Figure \ref{fig:lineCompare}, we show the relationship between emission line widths (FWHM) and luminosities. We find that no obvious correlations between the line width and the line luminosity. We further see that the \ha\ emission lines consistently show broad profiles with FWHM of several $1,000\textrm{ km s}^{-1}$. In contrast, the \oiii\ width distribution is concentrated around FWHM=$1,000-3,000\textrm{ km s}^{-1}$. For the 9 sources with both \ha\ and \oiii\ detections, the Kolmogorov-Smirnov (KS) test does not indicate a strong correlation between the line widths. This lack of correlation may suggest that the observed \ha\ and \oiii\ kinematics may originate from different mechanisms (e.g. BLR and NLR emission vs. outflows). To better understand the observed line kinematics, we compare our \oiii\ measurements with those of other quasar samples: $z<1$ Type 2 \citep{ZakamskaGreene2014}, $z\sim2$ Type 2 \citep{Harrison2012}, $1.5<z<3.5$ Type 1 \citep{Shen2016}, $2.3\lesssim z \lesssim 3.5$ quasars \citep{Bischetti2017}, and $z\sim2$ ERQs \citep{Perrotta2019MNRAS}. We can see that the \oiii\ emission lines in our sample are much fainter, yet are broad. The observed line properties are mostly consistent with other known Type-2 quasars. \cite{ZakamskaGreene2014} show median \oiii\ FWHM of $\sim890\textrm{ km s}^{-1}$, ranging between $340-4390\textrm{ km s}^{-1}$, after converting from $w_{80}$.

In addition to \ha, we see various low ionization lines like \oi$\lambda6300$\AA\ and \oii$\lambda3727$\AA\ in certain sources (e.g. J0213+0024, J2243+0017, J0149+0052, J0215+0042). \oi\ may be suggestive of shocks or outflows in the neutral medium. The non-detection of \feii\ further supports that these targets are indeed obscured or red quasars. 
The focus of this paper will be on the \ha\ and \hb$+$\oiii\ lines. 


\subsection{Rest-frame optical vs. IR}\label{sec:visIR}

Due to the nature of the color selection, all our targets have strong IR fluxes, despite variations in the optical emission. We compare the monochromatic IR luminosity at rest-frame $\lambda=5\micron$, which we denote as $L_{5\micron}\equiv\lambda L_{\lambda}(5\micron)$. 
Since the observed broadband \wise\ filters correspond to different rest-frame wavelengths, we must interpolate over the observed \wise\ fluxes, $F_{\lambda,\textrm{obs}}$. We assume the IR spectra take a power-law form $L_{\lambda}\propto\lambda^{\alpha}$ and interpolate to compute \LIR. We find an average slope of $\langle\alpha\rangle=1.05\pm0.6$ around $5\micron$. All sources are very IR luminous with $L_{5\micron}\sim 10^{46}-10^{47}\textrm{erg s}^{-1}$, comparable to the brightest known obscured quasars \citep[e.g.,][]{Ross2015MNRAS453}. 

We also calculate the bolometric luminosity by using the IR-bolometric correction $L_{\textrm{bol}}=8\times L_{3.45\micron}$ \citep{Hamann2017MNRAS,Perrotta2019MNRAS}. We denote $L_{3.45\micron}$ as the monochromatic luminosity at $\lambda=3.45\micron$, which is extrapolated from the computed \LIR\ value using the same piece-wise power-law interpolation described above. $L_{\textrm{bol}}$ reaches and exceeds $10^{47}\textrm{erg s}^{-1}$. We list the calculated \LIR\ and \Lbol\ values in Table \ref{tab:spectprop}. We calculate the source luminosities (emission line and bolometric) assuming isotropic emission. This assumption means that if obscuration makes the observed infrared emission anisotropic, as is the case with low redshift Type-2 quasars \citep{Liu2013MNRAS436}, then the actual \Lbol\ may even be higher as noted by \cite{ZakamskaGreene2014}.

In Figure \ref{fig:lineVwise} we show the \ha\ and \oiii\ line luminosities and kinematics, along with with the computed \LIR, which we use as a proxy for \Lbol. We do not see any obvious correlations between the \ha\ and \oiii\ line kinematics and \LIR. 

\subsection{Spectral energy distribution (SED) fitting}\label{sec:sed}

\begin{figure*}
	 \begin{center}
	 \begin{tabular}{ccc}
    \includegraphics[width=0.99\columnwidth]{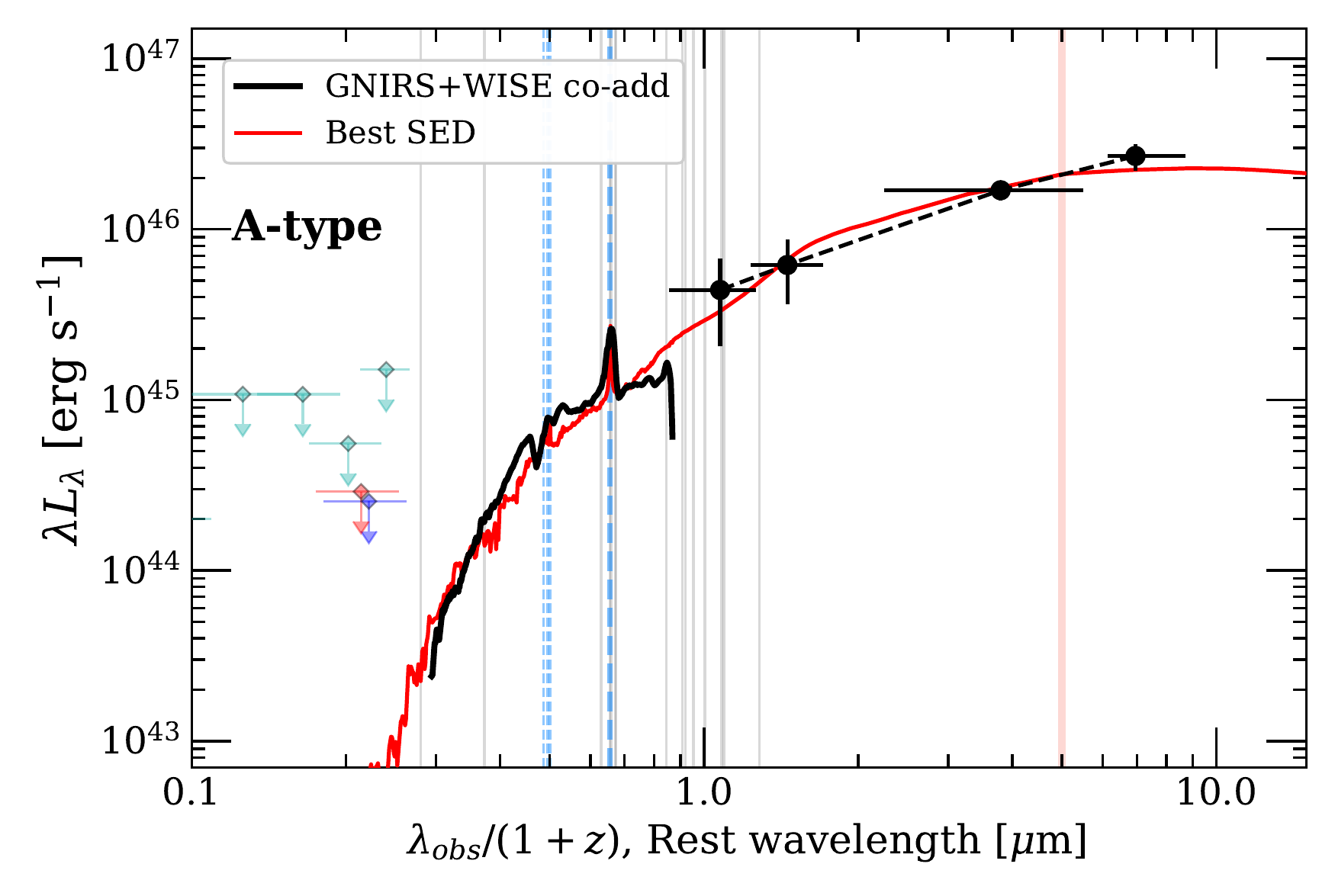}&
    \includegraphics[width=0.99\columnwidth]{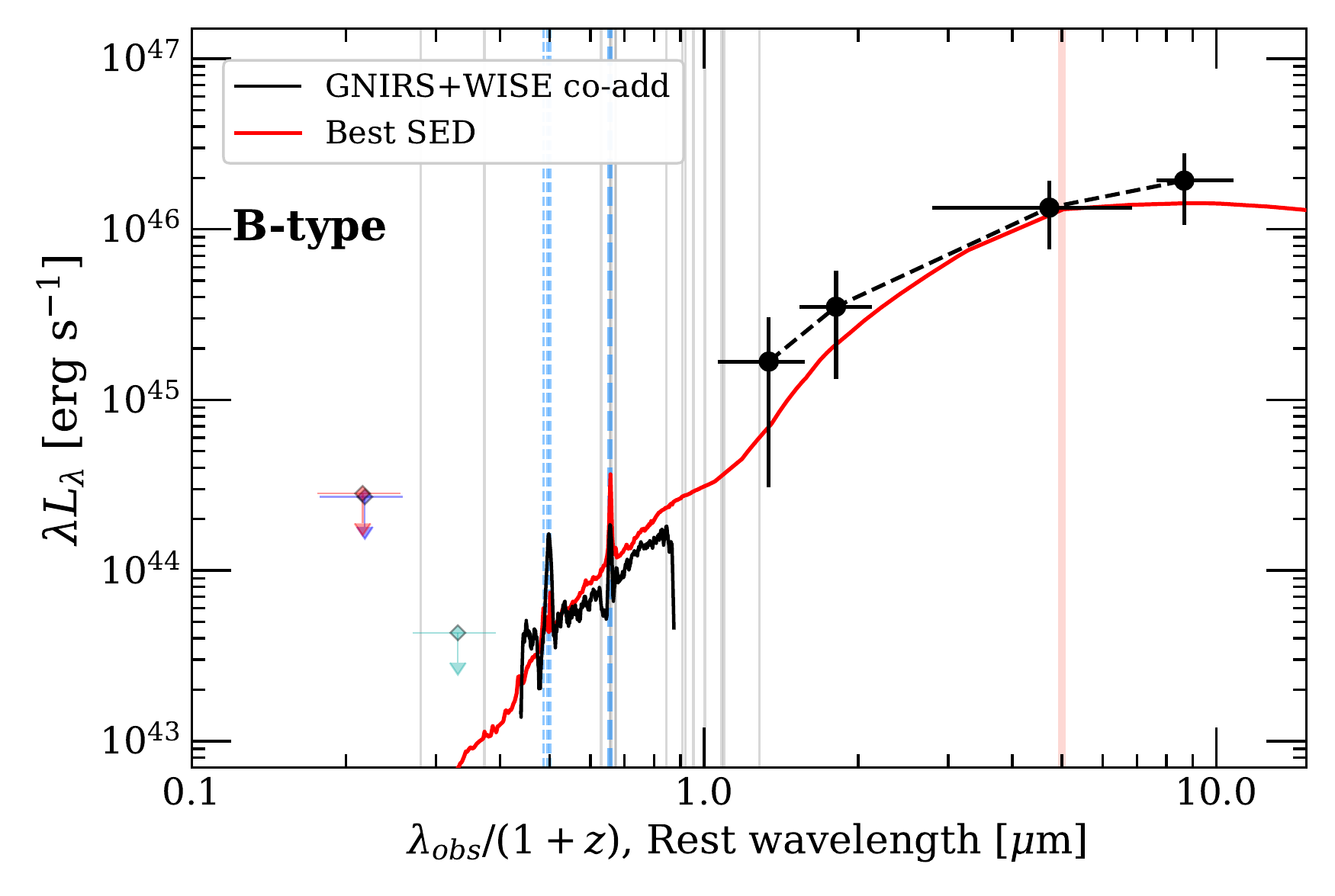}\\
    \includegraphics[width=0.99\columnwidth]{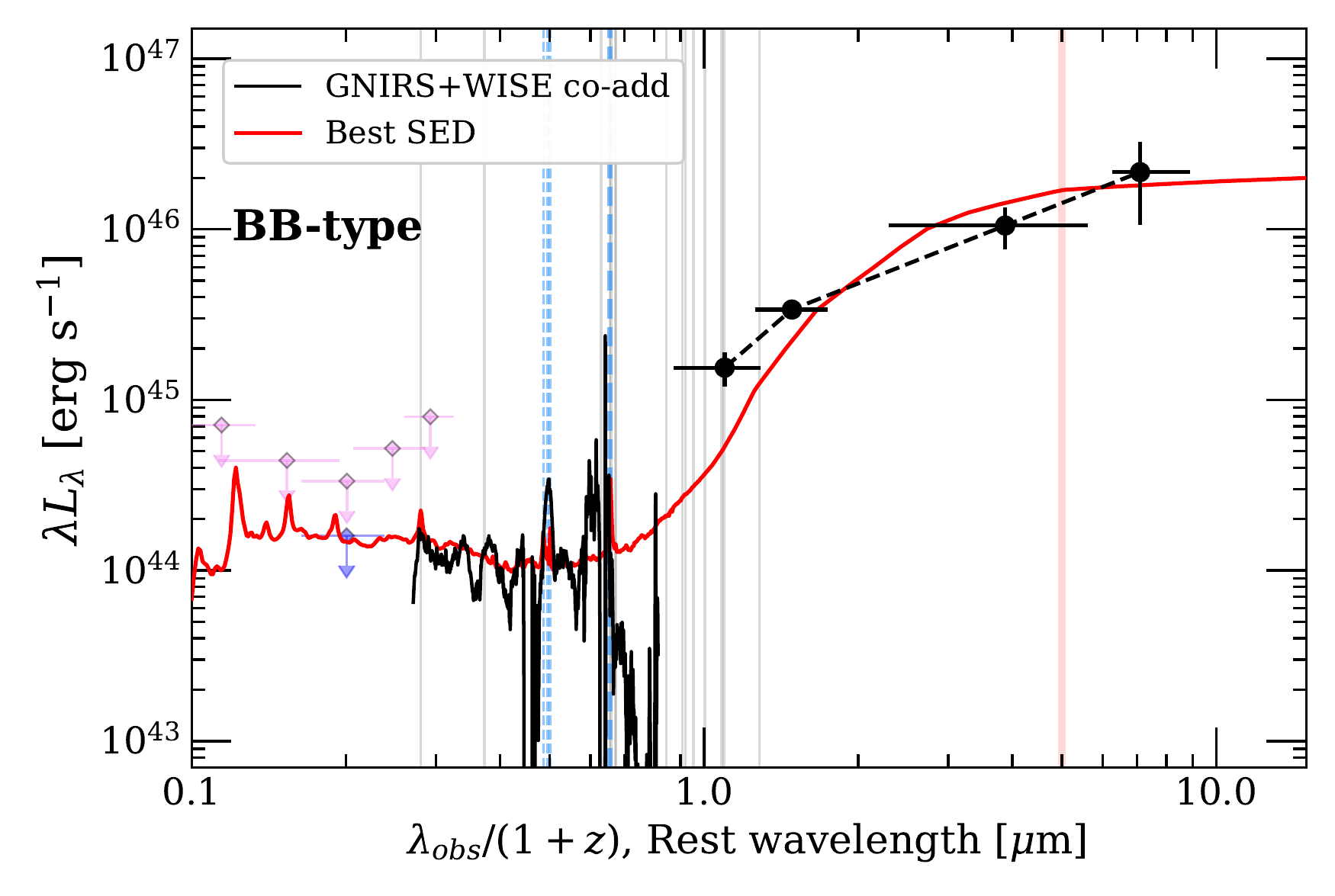}&
    \includegraphics[width=0.99\columnwidth]{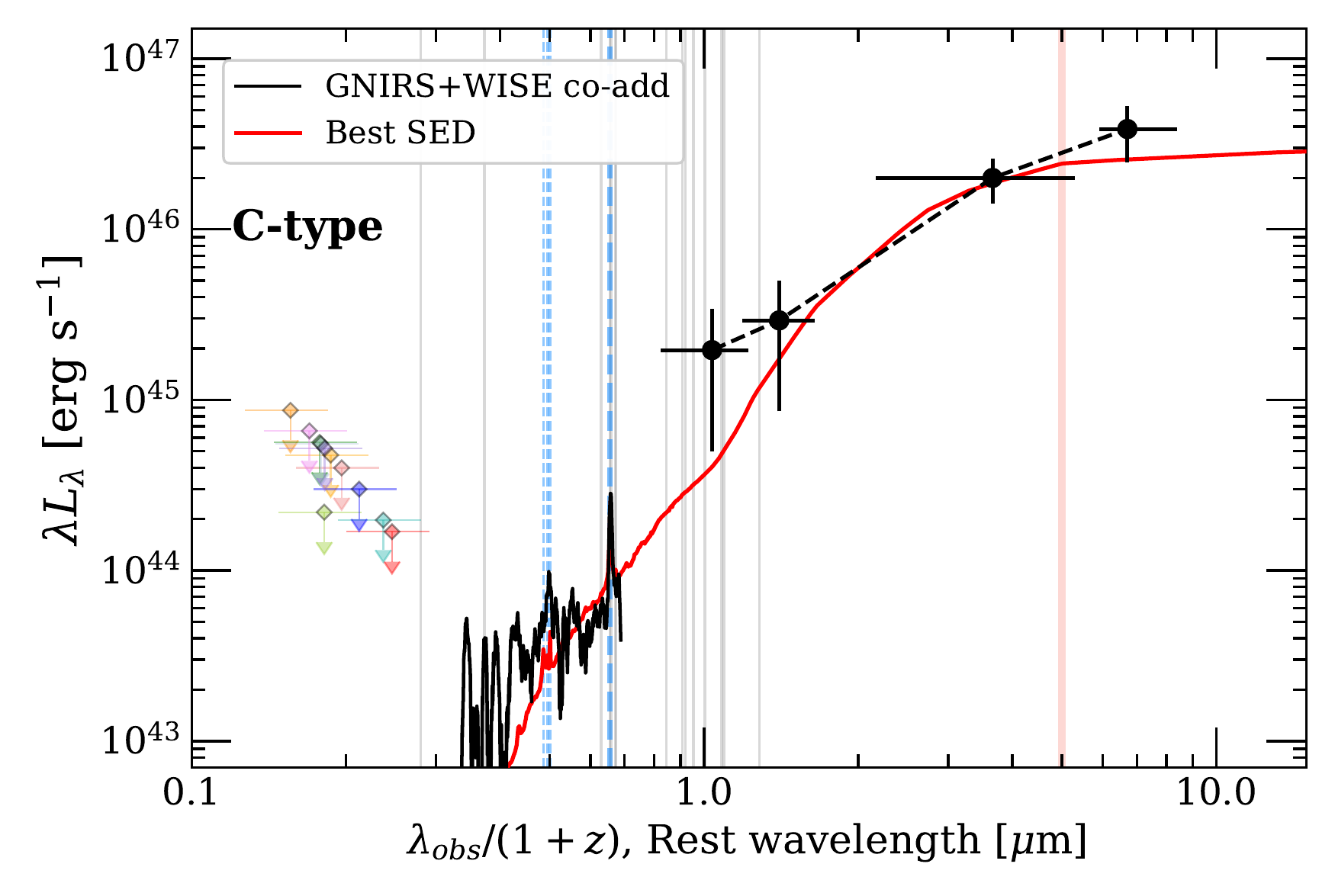}\\
	 \end{tabular}
	 \end{center}
	 \caption{We show the co-added \gnirs+\wise\ SEDs overplotted in solid black line and the best-fit SEDs in solid red line. (Top left) `A' strong continuum and emission lines;  (top right) `B' moderate continuum and emission lines;  (bottom left) `BB' blue-sloped continuum with moderate emission lines;  (bottom right) `C'+`D' weak continuum with single-line \ha\ detection. We indicate the \oiii\ and \ha\ emission lines (light blue dashed lines), the other low ionization lines (light grey lines), and \LIR\ (light red). We also plot the SDSS upper limits (light diamonds).} 
	 \label{fig:sed_types} 
\end{figure*}

\begin{figure}
	 \includegraphics[width=0.97\columnwidth]{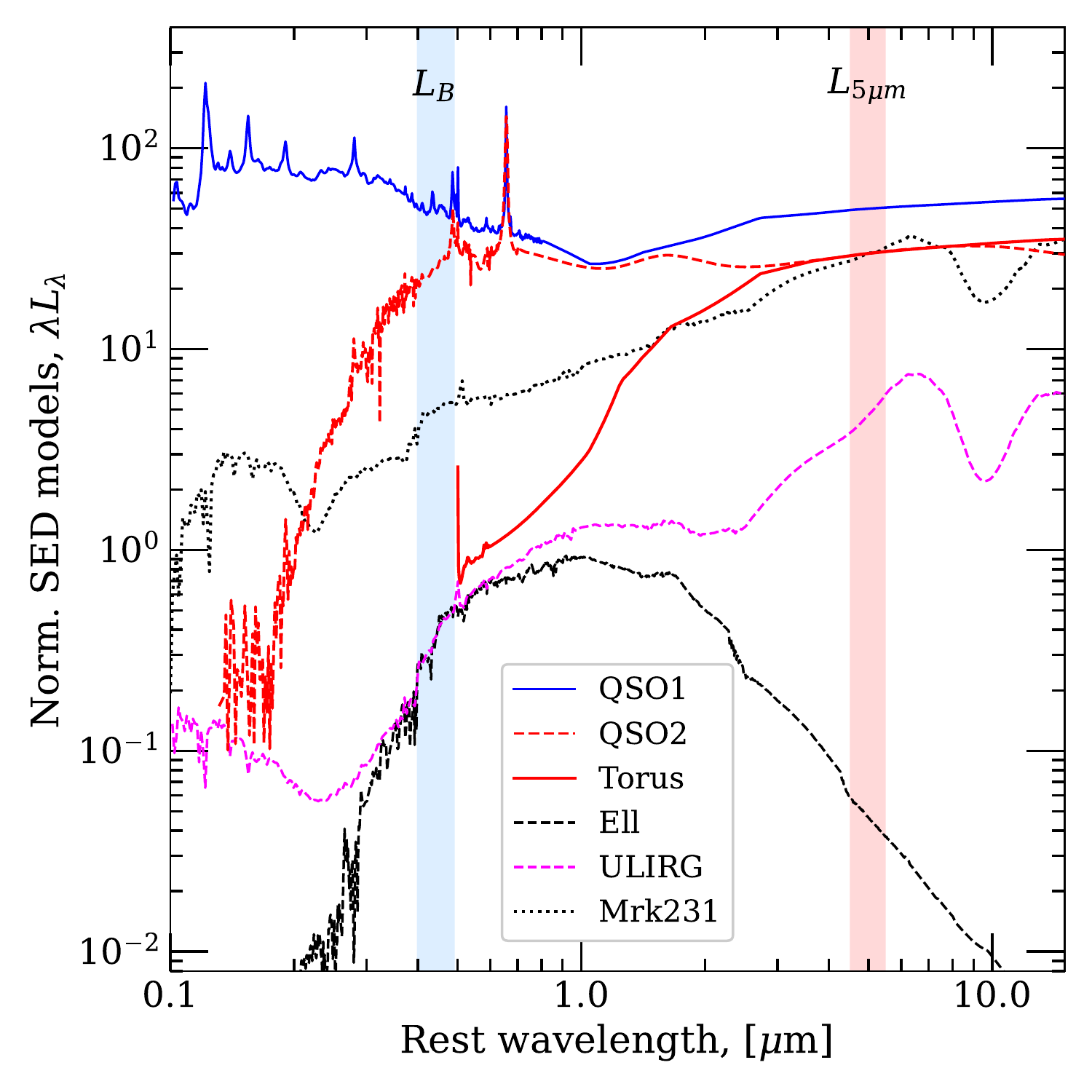}
	 \caption{The model SEDs from SWIRE \citep{Polletta2007ApJ}. The plots are normalized by $\lambda L_{\lambda}\sim10^{45}\textrm{ erg s}^{-1}$. The galaxy models are scaled at $L_B \sim10^{44.5}\textrm{ erg s}^{-1}$, and the quasar models are scaled at $L_{5\mu m} \sim10^{46.3}\textrm{ erg s}^{-1}$.}
	 \label{fig:sed_models} 
\end{figure}

Taking advantage of the optical and IR multi-wavelength data, we construct SEDs and perform SED fits to understand the nature of our objects. In Figure \ref{fig:sed_types} we show the SEDs separated by \Gemini/\gnirs\ spectral quality. Since many of the individual spectra suffer from low SNR, we co-add the \gnirs\ spectra according to the A/B/BB/C/D categories defined in Section \ref{sec:obsredux}. We show the co-added spectra in Figure \ref{fig:sed_types}. Interestingly, we find that the sources that did not show an obvious continuum individually now show a slight red slope after co-adding. We use the co-added spectra for the SED fits to simplify the analysis. 

Based on the bright \wise\ IR emission that reaches $10^{45.5}\textrm{-}10^{47}\textrm{erg s}^{-1}$, we can assume that the SED is most likely dominated by the quasar torus emission. It is unlikely for the MIR emission to originate from the host galaxy. For the brightest sources, if the observed light is dominated by the stellar light from the host galaxy, it would require B band luminosities of least $\lambda L_{\lambda}\sim10^{45}\textrm{erg s}^{-1}$ to produce $W4\sim10^{46}\textrm{erg s}^{-1}$, which is an order-of-magnitude larger than typical galaxies. For example, we know that ERQ host galaxies can be up to a few $10^{11} L_{\astrosun} \approx10^{44.5}\textrm{erg s}^{-1}$. Our sources are not more luminous than ERQs, so they are unlikely to live in even more massive galaxies, as discussed in Section \ref{sec:color}.

In Figure \ref{fig:sed_models} we show the different models considered. The MIR emission probed with \wise\ is consistent with a quasar+torus dominated spectrum with a heavy reddening. We account for these observed properties by constructing an SED model consisting of: quasar (Type 1 or 2) + torus + (host galaxy, if needed) with varying dust absorption. We use the SWIRE galaxy and quasar spectral templates \citep{Polletta2007ApJ}. We do not include the SDSS upper limits since they are too shallow to be constraining. Here, we determine the best SED (quasar/host/torus emission strength) and the estimated $A_V$ continuum reddening.

We assume a ``cold screen of dust absorption'' model with an extinction. For simplicity, we use the Milky Way extinction model \cite{Cardelli1989, Calzetti2000, Gordon2003}. We fit the coadded SEDs  following $F_{\textrm{obs}} = F_{\textrm{int}}e^{-\tau_{\lambda}}$ to determine the best $A_V$ extinction. We solve the following: 
\begin{equation}
    \log F_{\textrm{obs}} = \log k_1 F(\textrm{QSO+torus+gal}) - k_2 \tau_{\lambda}
\end{equation}
where $k_1$ is the normalization at $5$\mum\ for the template quasar$+$torus SED and $k_2$ is the normalization for the extinction curve $\tau_{\lambda}$. Initially, we attempted a simple single-phase absorption model; however, it was quickly apparent that a multi-component absorption model is necessary \citep[e.g.,][]{Ricci2017}, to describe the red-sloped ``bump'' seen with our GNIRS data. Here, we apply separate cold-dust absorption on each component (quasar, galaxy, and torus) and combine to produce a ``master'' SED. Both the quasar and torus components required heavy absorption, reaching $A_V\sim20$, whereas the galaxy component requires moderate absorption. We caution against over-interpreting the SEDs without detailed radiative transfer modeling. 

In Figure \ref{fig:sed_types}, we show the best SED fits. As expected, we find that the \wise-detected MIR emission is best fit with a torus model. The GNIRS-detected rest-frame optical spectra are best fit with a quasar spectrum reddened by at least $A_V\sim7-20$. The exception to this model is the blue excess `BB' targets, in which an additional non-reddened quasar spectrum is needed. We explore the blue excess in Sections \ref{sec:color} and \ref{sec:prop}. The brightest `A' sources require a bright galaxy component to explain the optical continuum. The exact properties of the host galaxy is unclear. What is clear is that the quasar spectrum dominates. We explore the interpretation of the reddening and the blue excess in the Discussion.

\begin{figure*}
	 \begin{center}
	 \begin{tabular}{c}
    \includegraphics[width=0.85\textwidth]{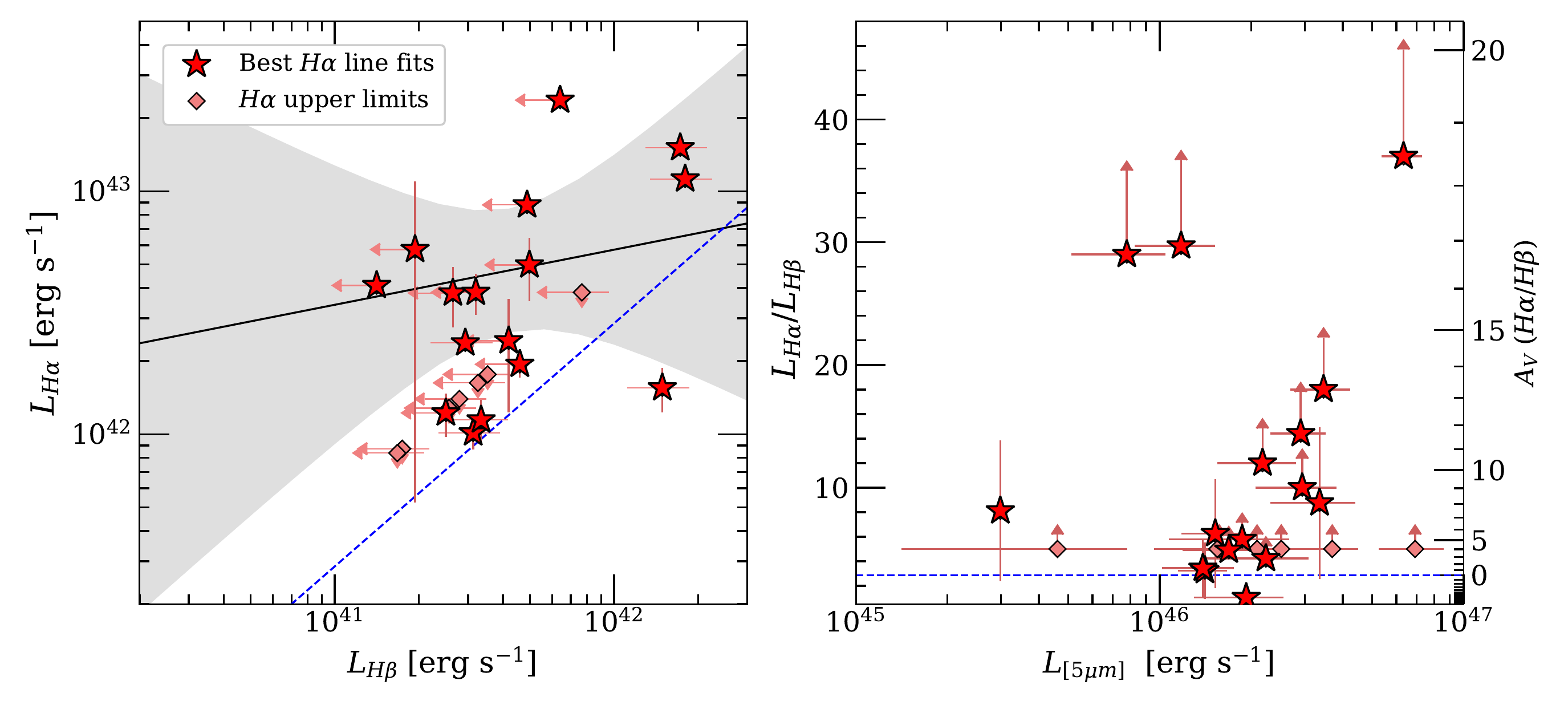}
	 \end{tabular}
	 \end{center}
	 \caption{(Left) Luminosities of \ha\ and \hb, compared with the theoretical expectation ${\rm H\alpha}/{\rm H\beta}\approx2.9$ shown with the blue dashed line. (Right) \HaHb\ Balmer decrement shown against \LIR. We also show the $A_V$ calculated with \HaHb. We see no obvious trends with \LIR. Since very few sources had sufficient \hb\ SNR, we place upper limits on $L_{H\beta}$, which translates to a lower bound on \HaHb.} 
	 \label{fig:balmerdec} 
\end{figure*}

\section{Discussion}
\label{sec:discuss}

\subsection{Efficiency of photometric selection}\label{sec:color}
The redshift distribution in Figure \ref{fig:zhist} shows that our selection preferentially selects $1<z<3$ objects. 
Unsurprisingly, the sources with stronger continuum and stronger emission lines have more reliable redshift determinations, as was noted by previous studies \citep[e.g.,][]{Lacy2013ApJS}. 

Past studies of W4- and optically-bright sources failed to uncover a high density of obscured quasars. In fact, the motivating color-selection of optically faint, yet red objects are similar to those of previous studies like ULIRGS,  HotDOGs, and ERQs (i.e. $r-W4>14$). Instead, in this study we take a step back by considering  IR-bright sources with $W4 \geq  5 \textrm{ mJy}$ that have more accurate positions from \spitz\ and that lack \sdss\ detections for a more robust selection. 

Figure \ref{fig:colors} serves as a diagnostic to compare the different color selection criteria for red and obscured quasars. The key differences in color-selection are as follows. HotDOGs primarily rely on W1 and W2 dropouts \citep{Eisenhardt2012ApJ}: $W1>17.4\textrm{ Vega mag}$ and either $(W4<7.7 \land W2-W4>8.2\textrm{ Vega mag})$ or $(W3<10.6 \land W2-W3>5.3 \textrm{ Vega mag})$, whereas we only require bright W4 detections. ERQs are defined by $i-W4>4.6$ colors and \civ\ equivalent width $>100$\AA\ \citep{Hamann2017MNRAS,Perrotta2019MNRAS}, whereas we strictly require weak $r$-band SDSS detections or non-detections in the SDSS optical photometry. \cite{Temple2019MNRAS} selection of reddened quasars applied a $W1-W2>0.85\textrm{ Vega mag}$ along with J/H/K-band photometric cuts, which covers a redder subset of the obscured quasars discussed here. This means that our selections identifies a reddened $1<z<3$ population that is both HotDOG-like (but bluer) and ERQ-like (but redder), as demonstrated in Figure \ref{fig:colors}.

Despite the similarities in the color-selection, our sources may constitute a distinct, yet similar population to HotDOGs and ERQs. For example, only 3 objects record any $i$-band SDSS fluxes, which means the ERQ $i-W4$ selection would fail to identify most of our sources. All of our sources have brighter \wise/W4 fluxes than ERQs. Since our sources probe a bluer selection than HotDOGs, we also see a greater range in the observed rest-frame optical luminosity, with 2 sources showing a distinct blue excess in their spectra. The blue continuum may be indicative of scattered light. The current optically-faint requirement selects against objects with a strong UV continuum, so naturally, we find very few blue-sloped targets. However, the detection of the `BB' objects may display an analogous component to those of ERQs and BHDs. We explore more in Section \ref{sec:prop}. 

Failure modes of the color-selected targets would result in either low redshift contaminants or featureless quasar spectra. However, low redshift contaminants should be minimized by limiting to optically-faint SDSS dropouts. Astrometry checks with \wise+\spitz\ ensures that the \Gemini/\gnirs\ pointing is robust. Barring heavy telluric contamination that prevent us from seeing the faint emission, the detection of featureless spectrum may simply reflect heavy obscuration of the inherent quasar spectrum, whether they are at circumnuclear or galactic scales. We explore more in Section \ref{sec:obsc}.


\subsection{Levels of obscuration}\label{sec:obsc}

As evident from the optical vs. IR luminosity in Figure \ref{fig:sed_types}, the quasars in our sample have substantial reddening. Here we characterize the origin of this reddening. In Figure \ref{fig:balmerdec}, we compare the \HaHb\ ratio to characterize the Balmer decrement. Photoionization by the quasar is likely the dominant mechanism that gives rise to Balmer emission in our targets. The balance of photoionization and recombination sets the ratios of Balmer lines at particular values -- from Case B recombination of hydrogen, the lower theoretical bound on ${\rm H\alpha}/{\rm H\beta}$ is $\approx2.98$ in dust-free gas at $T_e=10^4\textrm{ K}$ \citep{ Dopita2003,OsterbrockFerland2006}. Indeed, in unobscured Type-1 quasars, this value peaks at 3.3 \citep{kim2006}. Strong enhancement in this ratio is likely due to dust extinction. We estimate $A_V$ from the \HaHb\ ratio following \cite{Riffel2021}: 
\begin{equation}
    A_V = 7.22 \log\bigg(\frac{F_{H\alpha}/F_{H\beta}}{2.86}\bigg)
\end{equation}
assuming a $R_V=3.1$ extinction \citep{Cardelli1989}. Balmer decrement is typically a reliable technique to estimate dust extinction in galaxies \citep[e.g.,][]{Kennicutt1992}. However, this may not be the case if there is no reliable measure of extinction due to the low SNR of \hb\ \citep{Assef2015,Wu2018}. We see an excess of \ha\ of up to $\sim40$, which translates to $A_V\sim20\textrm{ mag}$. The observed Balmer decrement is consistent with the continuum reddening in Section \ref{sec:sed}. However, this is a little misleading since \hb\ is not detected for all sources, as indicated in Table \ref{tab:spectprop}. Sources with significant \hb\ emission have a median Balmer decrement of ${\rm H\alpha}/{\rm H\beta}\sim 3.5$, which correspond to $\langle A_V \rangle\sim4.8\pm2.8\textrm{ mag}$. For sources with no \hb\ detection, we indicate \HaHb\ as lower limits. We do not see any correlation with the IR luminosity. 

Although we find extreme reddening in some of our sources with large \HaHb\ values, we find that the average \HaHb\ reddening is consistent with other Type-2 quasars. A handful show much larger \HaHb\ than those of other known reddened quasars \citep[e.g.,][]{Glikman2018, Jun2020ApJ}. 
HotDOGs also show large $H\alpha/H\beta\gtrsim4$ values. 

We combine the cold dust extinction SED fits from Section \ref{sec:sed} and the Balmer decrement estimates. From the SED fits, these sources are likely quasar-dominated, with a bright MIR component from the reprocessed emission by the torus and a significantly reddened optical component. Considering the shape of the MIR SED, which is due to warm dust, and the large continuum reddening of the intrinsic quasar spectrum, circumnuclear obscuration is the most probable scenario. 
Although we cannot completely rule out the possibility of galactic scale obscuration, as \cite{Temple2019MNRAS} found using forbidden \oiii\ lines produced on large scales, our sources would nonetheless still require a circumnuclear torus to produce the observed MIR emission. While not impossible to have both circumnuclear and galactic scale obscuration, it is contrived. Our data do not have strong evidence in favor of galactic obscuration. 


\subsection{Properties of sources}\label{sec:prop}
As discussed earlier, the \gnirs\ observations revealed 3 types of sources with varying levels of continuum strengths: strong, moderate, and weak with weak lines. In Figure \ref{fig:sed_types} we show the representative SED plots. Despite differences in their rest-frame optical fluxes by up to 1.5 dexs, the IR fluxes show very similar values. The shapes of their IR SEDs are also similar. 
The IR luminosities reach up to $10^{46}-10^{47}\textrm{ erg s}^{-1}$. The best-fit SEDs suggest that these are likely due to the quasar torus emission, suggesting the dusty nature of these quasars, complementing the non-detections in the rest-frame UV and the broad \ha\ and \hb\ emission lines observed. These observations suggest that photometric color selection is an excellent metric for identifying obscured quasars at $z\sim2$. 

The typical line widths of the \oiii\ emission in our sources are $\sim1800\textrm{ km s}^{-1}$ at FWHM. \oiii\ is a forbidden emission line that must originate in low-density regions, typically well outside of the sphere of influence of the black hole. These velocity widths are too high to be contained by a realistic galaxy potential and are therefore indicative of likely galactic wind activity. However, colors alone are insufficient for identifying obscured quasars with extreme kinematics. ERQs show \oiii\ velocity widths in the range $1500-7400\textrm{ km s}^{-1}$ at FWHM, which are significantly higher than in the population presented here (Figure \ref{fig:lineVwise}, bottom right). ERQs were identified both by colors and by line diagnostics like \civ\ equivalent widths \citep{Hamann2017MNRAS,Perrotta2019MNRAS,Zakamska2019}. Using additional line diagnostics may be necessary to identify sources with potential quasar-driven galactic winds. We explore more in Section \ref{sec:typ2}.

Although black hole mass scaling relations from reverberation mapping (e.g. using continuum emission and \hb\ widths; \citealt{GreeneHo2005,Vestergaard2006}) are commonly used to estimate \Mbh, the nature of the obscuration complicates the comparisons to obscured quasars, as noted by \cite{Zakamska2019}. Supermassive black holes that are heavily obscured, have powerful winds, or are in the super-Eddington accretion limit do not necessarily obey the scaling relations because the inner broad-line region and the accretion disk are either not visible or are not in the same dynamical equilibrium as the sources used in the scaling relations. Since we have extremely low \hb\ SNR and very little optical continuum, we cannot obtain reliable black hole mass measurements. 
With these caveats in mind, we assume that the quasars are hosted by $M_{\textrm{BH}}\sim10^9 M_{\astrosun}$ black holes. 
This mass corresponds to an Eddington-limited ratio of $\lambda_{\textrm{Edd}}=L_{\textrm{bol}}/L_{\textrm{Edd}}\sim0.7$, where $L_{\textrm{Edd}}\approx1.3\times10^{47}\textrm{ erg s}^{-1}$. However, if \Lbol\ is underestimated due to the obscuration, it may place these quasars closer to the Eddington limited regime. And if the black holes are instead significantly more massive ($\sim10^{10} M_{\astrosun}$), they may be in the accretion regime close to that of normal quasars.

Another question is the origin of the strong MIR emission. For high redshift sources, we can be confident that these are quasar dominated. However, dusty, star-forming galaxies may also produce large MIR luminosities.  In Figure \ref{fig:sed_models}, we compare the different SED templates, including Type-1 QSO, Type-2 QSO, Torus, Starbursts, and ULIRGs. Fortunately, dusty star-forming galaxies like ULIRGs typically have cooler SEDs compared to the quasar torus emission. Also, considering the extreme optical faintness of our objects, it is unlikely to simultaneously produce high IR emission with low optical emission. As explored in Section \ref{sec:sed}, a more likely scenario is dominant torus emission with a faint host galaxy with $L_B\sim10^{44.5}\textrm{ erg s}^{-1}$ with extreme reddening of the UV/optical quasar spectrum. 

There are curious `BB' sources like J2334+0031 and J0041-0029 that display a noticeable blue continuum. These 2 sources also recorded SDSS fluxes at the survey's limiting magnitude, which is consistent with a UV quasar spectrum (Figure \ref{fig:sed_types}). A blue excess in obscured quasars is observed when light from the quasar continuum is scattered off the surrounding gas into our line of sight. This phenomenon has been directly confirmed in ERQs with spectropolarimetric observations \citep{Alexandroff2018}.  \citet{Zakamska2006, Alexandroff2018, Assef2022ApJ} have shown that polarization and blue excess reveal scattering from dust within the ionization cone. \cite{Assef2016, Assef2022ApJ} also identified a subset of the HotDOG population: so-called ``Blue HotDOGs'' (BHDs) with a blue/UV excess observed with SDSS photometry. Although we cannot ascertain the origin of the blue slope with the current \Gemini/\gnirs\ data, scattered light is a promising possibility and would strengthen the case that these sources are indeed obscured quasars. The discovery of these blue-slope sources and other obscured quasar candidates reveals that the population of obscured quasars has diverse properties. Also, these detections show that caution is needed when applying strict UV/optical dropout requirements to identify obscured quasars, further supporting our hypothesis that obscured quasars are simply missed. Using physically motivated, strategic color-selections, such as ours, can uncover the diverse populations of obscured quasars. 


\subsection{Comparing/linking with other obscured quasars}\label{sec:typ2}
As discussed earlier, the colors of our targets overlap with other types of obscured quasars. However, we see some key spectral differences, which may probe different physics. Perhaps we are probing the transitional link between the two populations. \citet{Assef2022ApJ} provided a physical model suggesting that the transitional link between HotDOGs, BHDs, and ERQs may be the interaction between the quasar-driven outflow and the obscuring medium. In this model, (1) the central quasar of HotDOGs is completely enshrouded by the obscuring material; (2) then massive, fast-moving outflows pierce through and ``blow out'' the obscuration as observed with ERQs (and possibly BHDs); and (3) finally, the column density of the obscuration is lowered, which is observed as heavily reddened Type-1 quasars. %

In the standard unification model of active nuclei, we expect the kinematics of \oiii\ to be similar in obscured and unobscured sources, since this line is produced on extended scales and reflects the kinematics of the gas in the host galaxy (or at least not in the sphere of influence of the black hole due to its low critical density). Indeed, \cite{Temple2019MNRAS} find that there are no statistical differences in the \oiii\ kinematics between obscured and unobscured quasars. One exceptional population of obscured quasars is ERQs, which show extremely high widths of \oiii, unmatched by any other quasar population \citep{Zakamska2016b, Perrotta2019MNRAS}. Our sample shows a broad range of \oiii\ velocities, similar to those of typical powerful quasars and somewhat below those in ERQs, as seen in Figures \ref{fig:lineCompare} and \ref{fig:lineVwise}. We do find 3 sources (J0024-0012, J0215+0042, and J2334+0031) reaching ERQ-like velocities and IR luminosities, as shown in Figure \ref{fig:lineVwise}. 

Here, we suggest that the transitional link may be observed in the \oiii\ luminosity and velocity vs. \LIR\ space, as shown in Figures \ref{fig:lineCompare} and  \ref{fig:lineVwise}. An interesting observation is that the ERQ selection requires a high \civ\ equivalent width \citep{Hamann2017MNRAS}, which may preferentially select reddened quasars with extreme \oiii\ emission and winds. Our \oiii\ emission lines (luminosity and kinematics) are nearly $\times10$ weaker than those seen in ERQs, yet likely significantly larger than HotDOGs. Since our targets show weak \oiii, it is difficult to determine if winds are present, although they are unlikely to have extreme winds like ERQs. According to the \cite{Assef2022ApJ} model, our selections may identify sources in the transition between stage (1), appearing bluer than HotDOGs, and stage (2), appearing redder than ERQs with weaker \oiii. By adjusting the detection criteria of optical and IR emission and nuclear line kinematics criteria, we may be able to recover more obscured populations.


If our targets are indeed obscured quasars, then a literature review for spectral data on our broader samples of 373 objects suggests that the number density of obscured quasars is between $0.17-0.93\textrm{ deg}^2$. This density is potentially comparable to the confirmed unobscured density to the same depth in W4 but does not exceed the unobscured number density. A conservative interpretation is that unobscured quasars may only outnumber obscured quasars by a small factor. However, there is a caveat that this selection is biased toward bright $W4<8\textrm{ Vega mag}$ sources, which may fail to identify fainter quasars. Although more comprehensive studies are necessary to resolve the number density estimates, our spectroscopic confirmation of these targets as obscured quasars support our hypothesis that obscured quasars have been missed by previous studies. 





\section{Conclusions} 
\label{sec:concl}
The census of obscured quasars at high redshift remains incomplete. We present \Gemini/\gnirs\ spectra of 24 luminous, red quasar candidates at $z\sim2$ selected using \wise/W4 photometry from the \sdss\ Stripe 82 field. These sources have little to no detectable optical flux, yet have high \wise/W4 fluxes ($\textrm{W4}\geq5\textrm{ mJy}$). This selection aims to probe the peak of the quasar-heated dust emission at rest-frame $\sim10\micron$ while requiring high optical obscuration with SDSS dropouts. The use of the \spitz\ data ensures precise astrometry for these optically faint sources. This study probes a slightly different, yet complementary selection regime compared to that of other known red, obscured quasars like HotDOGs and ERQs. 

From spectroscopic analysis, we find a diverse range of spectral types. Two sources show strong continuum and strong emission lines, six show moderate continuum and emission lines, and others show no continuum with weak emission lines if detected. All sources with continuum detection have a red slope. We detect moderately strong \ha\ and \oiii\ emission lines, which we use to determine the redshift of these sources. We find that our \wise-selected IR bright color-selection method selects $1<z<3$ objects with the average redshift of $\langle z_{\textrm{best}}\rangle= 2.008\pm0.45$. 

Due to the low SNR of the emission lines detected, we fit single Gaussians to the continuum-subtracted spectra and we cannot study the details of line kinematics. From these fits, we find broad \ha\ emission lines with FWHM ranging several $10^3\textrm{ km s}^{-1}$ and luminosities reaching $10^{42}-10^{43.4} \textrm{erg s}^{-1}$. Very little \hb\ is detected, whereas the \oiii\ doublet show clear detections. \oiii$\lambda5009$\AA\ show moderate line strengths with luminosities $10^{42}-10^{43} \textrm{erg s}^{-1}$ and FWHM of several $10^3\textrm{ km s}^{-1}$. 

The red slope and the lack of \hb\ detections with large \HaHb\ ratios are indicative that these sources are experiencing significant extinction with $\langle A_V (H\alpha/H\beta) \rangle\sim4.8\pm2.8\textrm{ mag}$. We estimate the $A_V$ extinction from the SED to probe the origin of the observed obscuration. We find typical values of $A_{V,\textrm{SED}}\sim7-20$. The SED modeling indicate that these are indeed obscured quasars with a heavily-reddened quasar+torus dominated spectrum.

As expected from the large \wise/W4 fluxes, we find that these sources have large \Lbol\ reaching $\sim10^{46}-10^{47} \textrm{erg s}^{-1}$, which is comparable to some of the brightest known quasars. Compared to known unobscured and obscured quasars at the similar redshift and \Lbol, we find that the \oiii\ line widths of our sources are similar to Type-1 quasars. Our sources do not display any signs of powerful quasar-driven winds such as those seen in ERQs \citep{Zakamska2016b, Perrotta2019MNRAS}. This suggests that the ``red-ness'' of the quasar does not necessarily ensure the presence of strong winds. 

The population of quasars we have discovered and characterized in this paper appears to be an intermediate one between HotDOGs and ERQs, with HotDOGs likely the most obscured. The differences in the properties are apparent from the shapes of the optical-to-IR SED and from the relative strengths of the emission lines. The red quasars of \cite{Temple2019MNRAS} are significantly less obscured. It would be interesting to determine whether the column densities of X-ray absorbing gas in comparison to those of ERQs \citep{Goulding2018ApJ} and HotDOGs \citep{Vito2018} support this picture.

\section*{Acknowledgements}
\label{sec:acknow}


This work was enabled by observations made from the Gemini North telescope, located within the Maunakea Science Reserve and adjacent to the summit of Maunakea. We are grateful for the privilege of observing the Universe from a place that is unique in both its astronomical quality and its cultural significance.

Based on observations obtained at the international Gemini Observatory, a program of NSF’s NOIRLab, which is managed by the Association of Universities for Research in Astronomy (AURA) under a cooperative agreement with the National Science Foundation. on behalf of the Gemini Observatory partnership: the National Science Foundation (United States), National Research Council (Canada), Agencia Nacional de Investigaci\'{o}n y Desarrollo (Chile), Ministerio de Ciencia, Tecnolog\'{i}a e Innovaci\'{o}n (Argentina), Minist\'{e}rio da Ci\^{e}ncia, Tecnologia, Inova\c{c}\~{o}es e Comunica\c{c}\~{o}es (Brazil), and Korea Astronomy and Space Science Institute (Republic of Korea).

This research made use of \pypeit,\footnote{\url{https://pypeit.readthedocs.io/en/latest/}}
a Python package for semi-automated reduction of astronomical slit-based spectroscopy
\citep{pypeit:joss_pub, pypeit:zenodo}. The authors would especially like to thank the \pypeit\ team members for their assistance with running the pipeline. This work also made use of \texttt{Astropy}\footnote{\url{http://www.astropy.org}} \citep{astropy2022}.

Y.I. acknowledges support by Space@Hopkins Graduate Fellowship. N.L.Z. is supported at the IAS by the J. Robert Oppenheimer Visiting Professorship and the Bershadsky Fund. N.L.Z and G.T.R. acknowledge support of NASA ADAP NNX12AI49G and NNX17AF04G. The authors also thank Jan-Torge Schindler and Riccardo Nanni for their contributions.


\section*{Data Availability} \label{sec:dataAvail}
Data are publicly available from the Gemini Observatory Archive (GOA) located at \url{https://archive.gemini.edu} under \progID\ (PI: Richards).



\bibliographystyle{mnras}
\bibliography{zerqBib} 

\begin{thebibliography}{}
\makeatletter
\relax
\def\mn@urlcharsother{\let\do\@makeother \do\$\do\&\do\#\do\^\do\_\do\%\do\~}
\def\mn@doi{\begingroup\mn@urlcharsother \@ifnextchar [ {\mn@doi@}
  {\mn@doi@[]}}
\def\mn@doi@[#1]#2{\def\@tempa{#1}\ifx\@tempa\@empty \href
  {http://dx.doi.org/#2} {doi:#2}\else \href {http://dx.doi.org/#2} {#1}\fi
  \endgroup}
\def\mn@eprint#1#2{\mn@eprint@#1:#2::\@nil}
\def\mn@eprint@arXiv#1{\href {http://arxiv.org/abs/#1} {{\tt arXiv:#1}}}
\def\mn@eprint@dblp#1{\href {http://dblp.uni-trier.de/rec/bibtex/#1.xml}
  {dblp:#1}}
\def\mn@eprint@#1:#2:#3:#4\@nil{\def\@tempa {#1}\def\@tempb {#2}\def\@tempc
  {#3}\ifx \@tempc \@empty \let \@tempc \@tempb \let \@tempb \@tempa \fi \ifx
  \@tempb \@empty \def\@tempb {arXiv}\fi \@ifundefined
  {mn@eprint@\@tempb}{\@tempb:\@tempc}{\expandafter \expandafter \csname
  mn@eprint@\@tempb\endcsname \expandafter{\@tempc}}}

\bibitem[\protect\citeauthoryear{{Alexandroff} et~al.,}{{Alexandroff}
  et~al.}{2018}]{Alexandroff2018}
{Alexandroff} R.~M.,  et~al., 2018, \mn@doi [\mnras] {10.1093/mnras/sty1685},
  \href {https://ui.adsabs.harvard.edu/abs/2018MNRAS.479.4936A} {479, 4936}

\bibitem[\protect\citeauthoryear{{Antonucci}}{{Antonucci}}{1993}]{Antonucci1993ARAA}
{Antonucci} R.,  1993, \mn@doi [\araa] {10.1146/annurev.aa.31.090193.002353},
  \href {https://ui.adsabs.harvard.edu/abs/1993ARA&A..31..473A} {31, 473}

\bibitem[\protect\citeauthoryear{{Assef} et~al.,}{{Assef}
  et~al.}{2015}]{Assef2015}
{Assef} R.~J.,  et~al., 2015, \mn@doi [\apj] {10.1088/0004-637X/804/1/27},
  \href {https://ui.adsabs.harvard.edu/abs/2015ApJ...804...27A} {804, 27}

\bibitem[\protect\citeauthoryear{{Assef} et~al.,}{{Assef}
  et~al.}{2016}]{Assef2016}
{Assef} R.~J.,  et~al., 2016, \mn@doi [\apj] {10.3847/0004-637X/819/2/111},
  \href {https://ui.adsabs.harvard.edu/abs/2016ApJ...819..111A} {819, 111}

\bibitem[\protect\citeauthoryear{{Assef} et~al.,}{{Assef}
  et~al.}{2022}]{Assef2022ApJ}
{Assef} R.~J.,  et~al., 2022, \mn@doi [\apj] {10.3847/1538-4357/ac77fc}, \href
  {https://ui.adsabs.harvard.edu/abs/2022ApJ...934..101A} {934, 101}

\bibitem[\protect\citeauthoryear{{Astropy Collaboration} et~al.,}{{Astropy
  Collaboration} et~al.}{2022}]{astropy2022}
{Astropy Collaboration} et~al., 2022, \mn@doi [\apj]
  {10.3847/1538-4357/ac7c74}, \href
  {https://ui.adsabs.harvard.edu/abs/2022ApJ...935..167A} {935, 167}

\bibitem[\protect\citeauthoryear{{Banerji}, {Alaghband-Zadeh}, {Hewett}  \&
  {McMahon}}{{Banerji} et~al.}{2015}]{Banerji2015}
{Banerji} M.,  {Alaghband-Zadeh} S.,  {Hewett} P.~C.,   {McMahon} R.~G.,  2015,
  \mn@doi [\mnras] {10.1093/mnras/stu2649}, \href
  {https://ui.adsabs.harvard.edu/abs/2015MNRAS.447.3368B} {447, 3368}

\bibitem[\protect\citeauthoryear{{Bischetti} et~al.,}{{Bischetti}
  et~al.}{2017}]{Bischetti2017}
{Bischetti} M.,  et~al., 2017, \mn@doi [\aap] {10.1051/0004-6361/201629301},
  \href {https://ui.adsabs.harvard.edu/abs/2017A&A...598A.122B} {598, A122}

\bibitem[\protect\citeauthoryear{{Brand} et~al.,}{{Brand}
  et~al.}{2007}]{Brand2007}
{Brand} K.,  et~al., 2007, \mn@doi [\apj] {10.1086/518119}, \href
  {https://ui.adsabs.harvard.edu/abs/2007ApJ...663..204B} {663, 204}

\bibitem[\protect\citeauthoryear{{Calzetti}, {Armus}, {Bohlin}, {Kinney},
  {Koornneef}  \& {Storchi-Bergmann}}{{Calzetti} et~al.}{2000}]{Calzetti2000}
{Calzetti} D.,  {Armus} L.,  {Bohlin} R.~C.,  {Kinney} A.~L.,  {Koornneef} J.,
   {Storchi-Bergmann} T.,  2000, \mn@doi [\apj] {10.1086/308692}, \href
  {https://ui.adsabs.harvard.edu/abs/2000ApJ...533..682C} {533, 682}

\bibitem[\protect\citeauthoryear{{Canalizo} \& {Stockton}}{{Canalizo} \&
  {Stockton}}{2001}]{CanalizoStockton2001}
{Canalizo} G.,  {Stockton} A.,  2001, \mn@doi [\apj] {10.1086/321520}, \href
  {https://ui.adsabs.harvard.edu/abs/2001ApJ...555..719C} {555, 719}

\bibitem[\protect\citeauthoryear{{Cardelli}, {Clayton}  \& {Mathis}}{{Cardelli}
  et~al.}{1989}]{Cardelli1989}
{Cardelli} J.~A.,  {Clayton} G.~C.,   {Mathis} J.~S.,  1989, \mn@doi [\apj]
  {10.1086/167900}, \href
  {https://ui.adsabs.harvard.edu/abs/1989ApJ...345..245C} {345, 245}

\bibitem[\protect\citeauthoryear{{Dey} et~al.,}{{Dey}
  et~al.}{2008}]{Dey2008ApJ}
{Dey} A.,  et~al., 2008, \mn@doi [\apj] {10.1086/529516}, \href
  {https://ui.adsabs.harvard.edu/abs/2008ApJ...677..943D} {677, 943}

\bibitem[\protect\citeauthoryear{{Donley} et~al.,}{{Donley}
  et~al.}{2012}]{Donley2012ApJ}
{Donley} J.~L.,  et~al., 2012, \mn@doi [\apj] {10.1088/0004-637X/748/2/142},
  \href {https://ui.adsabs.harvard.edu/abs/2012ApJ...748..142D} {748, 142}

\bibitem[\protect\citeauthoryear{{Dopita} \& {Sutherland}}{{Dopita} \&
  {Sutherland}}{2003}]{Dopita2003}
{Dopita} M.~A.,  {Sutherland} R.~S.,  2003, {Astrophysics of the diffuse
  universe}, \mn@doi{10.1007/978-3-662-05866-4.
}

\bibitem[\protect\citeauthoryear{{Eisenhardt} et~al.,}{{Eisenhardt}
  et~al.}{2012}]{Eisenhardt2012ApJ}
{Eisenhardt} P. R.~M.,  et~al., 2012, \mn@doi [\apj]
  {10.1088/0004-637X/755/2/173}, \href
  {https://ui.adsabs.harvard.edu/abs/2012ApJ...755..173E} {755, 173}

\bibitem[\protect\citeauthoryear{{Elias}, {Rodgers}, {Joyce}, {Lazo},
  {Doppmann}, {Winge}  \& {Rodr{\'\i}guez-Ardila}}{{Elias}
  et~al.}{2006a}]{Elias2006SPIE6269E14E}
{Elias} J.~H.,  {Rodgers} B.,  {Joyce} R.~R.,  {Lazo} M.,  {Doppmann} G.,
  {Winge} C.,   {Rodr{\'\i}guez-Ardila} A.,  2006a, in {McLean} I.~S.,  {Iye}
  M.,  eds,  Society of Photo-Optical Instrumentation Engineers (SPIE)
  Conference Series Vol. 6269, Society of Photo-Optical Instrumentation
  Engineers (SPIE) Conference Series. p. 626914, \mn@doi{10.1117/12.671765}

\bibitem[\protect\citeauthoryear{{Elias}, {Joyce}, {Liang}, {Muller}, {Hileman}
   \& {George}}{{Elias} et~al.}{2006b}]{Elias2006SPIE6269E4CE}
{Elias} J.~H.,  {Joyce} R.~R.,  {Liang} M.,  {Muller} G.~P.,  {Hileman} E.~A.,
   {George} J.~R.,  2006b, in {McLean} I.~S.,  {Iye} M.,  eds,  Society of
  Photo-Optical Instrumentation Engineers (SPIE) Conference Series Vol. 6269,
  Society of Photo-Optical Instrumentation Engineers (SPIE) Conference Series.
  p. 62694C, \mn@doi{10.1117/12.671817}

\bibitem[\protect\citeauthoryear{{Elitzur} \& {Shlosman}}{{Elitzur} \&
  {Shlosman}}{2006}]{Elitzur2006ApJ}
{Elitzur} M.,  {Shlosman} I.,  2006, \mn@doi [\apjl] {10.1086/508158}, \href
  {https://ui.adsabs.harvard.edu/abs/2006ApJ...648L.101E} {648, L101}

\bibitem[\protect\citeauthoryear{{Glikman} et~al.,}{{Glikman}
  et~al.}{2012}]{Glikman2012}
{Glikman} E.,  et~al., 2012, \mn@doi [\apj] {10.1088/0004-637X/757/1/51}, \href
  {https://ui.adsabs.harvard.edu/abs/2012ApJ...757...51G} {757, 51}

\bibitem[\protect\citeauthoryear{{Glikman} et~al.,}{{Glikman}
  et~al.}{2013}]{Glikman2013}
{Glikman} E.,  et~al., 2013, \mn@doi [\apj] {10.1088/0004-637X/778/2/127},
  \href {https://ui.adsabs.harvard.edu/abs/2013ApJ...778..127G} {778, 127}

\bibitem[\protect\citeauthoryear{{Glikman} et~al.,}{{Glikman}
  et~al.}{2018}]{Glikman2018}
{Glikman} E.,  et~al., 2018, \mn@doi [\apj] {10.3847/1538-4357/aac5d8}, \href
  {https://ui.adsabs.harvard.edu/abs/2018ApJ...861...37G} {861, 37}

\bibitem[\protect\citeauthoryear{{Gordon}, {Clayton}, {Misselt}, {Landolt}  \&
  {Wolff}}{{Gordon} et~al.}{2003}]{Gordon2003}
{Gordon} K.~D.,  {Clayton} G.~C.,  {Misselt} K.~A.,  {Landolt} A.~U.,   {Wolff}
  M.~J.,  2003, \mn@doi [\apj] {10.1086/376774}, \href
  {https://ui.adsabs.harvard.edu/abs/2003ApJ...594..279G} {594, 279}

\bibitem[\protect\citeauthoryear{{Goulding} et~al.,}{{Goulding}
  et~al.}{2018}]{Goulding2018ApJ}
{Goulding} A.~D.,  et~al., 2018, \mn@doi [\apj] {10.3847/1538-4357/aab040},
  \href {https://ui.adsabs.harvard.edu/abs/2018ApJ...856....4G} {856, 4}

\bibitem[\protect\citeauthoryear{{Greene} \& {Ho}}{{Greene} \&
  {Ho}}{2005}]{GreeneHo2005}
{Greene} J.~E.,  {Ho} L.~C.,  2005, \mn@doi [\apj] {10.1086/431897}, \href
  {https://ui.adsabs.harvard.edu/abs/2005ApJ...630..122G} {630, 122}

\bibitem[\protect\citeauthoryear{{Hamann} et~al.,}{{Hamann}
  et~al.}{2017}]{Hamann2017MNRAS}
{Hamann} F.,  et~al., 2017, \mn@doi [\mnras] {10.1093/mnras/stw2387}, \href
  {https://ui.adsabs.harvard.edu/abs/2017MNRAS.464.3431H} {464, 3431}

\bibitem[\protect\citeauthoryear{{Harrison} et~al.,}{{Harrison}
  et~al.}{2012}]{Harrison2012}
{Harrison} C.~M.,  et~al., 2012, \mn@doi [\mnras]
  {10.1111/j.1365-2966.2012.21723.x}, \href
  {https://ui.adsabs.harvard.edu/abs/2012MNRAS.426.1073H} {426, 1073}

\bibitem[\protect\citeauthoryear{{Hasinger}}{{Hasinger}}{2008}]{Hasinger2008AA}
{Hasinger} G.,  2008, \mn@doi [\aap] {10.1051/0004-6361:200809839}, \href
  {https://ui.adsabs.harvard.edu/abs/2008A&A...490..905H} {490, 905}

\bibitem[\protect\citeauthoryear{{Hennawi} \& {Prochaska}}{{Hennawi} \&
  {Prochaska}}{2013}]{Hennawi2013ApJ766}
{Hennawi} J.~F.,  {Prochaska} J.~X.,  2013, \mn@doi [\apj]
  {10.1088/0004-637X/766/1/58}, \href
  {https://ui.adsabs.harvard.edu/abs/2013ApJ...766...58H} {766, 58}

\bibitem[\protect\citeauthoryear{{Hickox} \& {Alexander}}{{Hickox} \&
  {Alexander}}{2018}]{HickoxAlexander2018}
{Hickox} R.~C.,  {Alexander} D.~M.,  2018, \mn@doi [\araa]
  {10.1146/annurev-astro-081817-051803}, \href
  {https://ui.adsabs.harvard.edu/abs/2018ARA&A..56..625H} {56, 625}

\bibitem[\protect\citeauthoryear{{Hopkins}, {Hernquist}, {Cox}, {Di Matteo},
  {Robertson}  \& {Springel}}{{Hopkins} et~al.}{2006}]{Hopkins2006}
{Hopkins} P.~F.,  {Hernquist} L.,  {Cox} T.~J.,  {Di Matteo} T.,  {Robertson}
  B.,   {Springel} V.,  2006, \mn@doi [\apjs] {10.1086/499298}, \href
  {https://ui.adsabs.harvard.edu/abs/2006ApJS..163....1H} {163, 1}

\bibitem[\protect\citeauthoryear{{Ishikawa}, {Goulding}, {Zakamska}, {Hamann},
  {Vayner}, {Veilleux}  \& {Wylezalek}}{{Ishikawa}
  et~al.}{2021}]{Ishikawa2021ERQ}
{Ishikawa} Y.,  {Goulding} A.~D.,  {Zakamska} N.~L.,  {Hamann} F.,  {Vayner}
  A.,  {Veilleux} S.,   {Wylezalek} D.,  2021, \mn@doi [\mnras]
  {10.1093/mnras/stab137}, \href
  {https://ui.adsabs.harvard.edu/abs/2021MNRAS.502.3769I} {502, 3769}

\bibitem[\protect\citeauthoryear{{Jun} et~al.,}{{Jun}
  et~al.}{2020}]{Jun2020ApJ}
{Jun} H.~D.,  et~al., 2020, \mn@doi [\apj] {10.3847/1538-4357/ab5e7b}, \href
  {https://ui.adsabs.harvard.edu/abs/2020ApJ...888..110J} {888, 110}

\bibitem[\protect\citeauthoryear{{Kennicutt}}{{Kennicutt}}{1992}]{Kennicutt1992}
{Kennicutt} Robert~C. J.,  1992, \mn@doi [\apj] {10.1086/171154}, \href
  {https://ui.adsabs.harvard.edu/abs/1992ApJ...388..310K} {388, 310}

\bibitem[\protect\citeauthoryear{{Kim}, {Ho}  \& {Im}}{{Kim}
  et~al.}{2006}]{kim2006}
{Kim} M.,  {Ho} L.~C.,   {Im} M.,  2006, \mn@doi [\apj] {10.1086/501422}, \href
  {https://ui.adsabs.harvard.edu/abs/2006ApJ...642..702K} {642, 702}

\bibitem[\protect\citeauthoryear{{Konigl} \& {Kartje}}{{Konigl} \&
  {Kartje}}{1994}]{KoniglKartje1994}
{Konigl} A.,  {Kartje} J.~F.,  1994, \mn@doi [\apj] {10.1086/174746}, \href
  {https://ui.adsabs.harvard.edu/abs/1994ApJ...434..446K} {434, 446}

\bibitem[\protect\citeauthoryear{{Krolik} \& {Begelman}}{{Krolik} \&
  {Begelman}}{1988}]{KrolikBegelman1988}
{Krolik} J.~H.,  {Begelman} M.~C.,  1988, \mn@doi [\apj] {10.1086/166414},
  \href {https://ui.adsabs.harvard.edu/abs/1988ApJ...329..702K} {329, 702}

\bibitem[\protect\citeauthoryear{{LaMassa} et~al.,}{{LaMassa}
  et~al.}{2015}]{LaMassa2015}
{LaMassa} S.~M.,  et~al., 2015, \mn@doi [\apj] {10.1088/0004-637X/800/2/144},
  \href {https://ui.adsabs.harvard.edu/abs/2015ApJ...800..144L} {800, 144}

\bibitem[\protect\citeauthoryear{{LaMassa}, {Yaqoob}, {Boorman}, {Tzanavaris},
  {Levenson}, {Gandhi}, {Ptak}  \& {Heckman}}{{LaMassa}
  et~al.}{2019}]{LaMassa2019ApJ}
{LaMassa} S.~M.,  {Yaqoob} T.,  {Boorman} P.~G.,  {Tzanavaris} P.,  {Levenson}
  N.~A.,  {Gandhi} P.,  {Ptak} A.~F.,   {Heckman} T.~M.,  2019, \mn@doi [\apj]
  {10.3847/1538-4357/ab552c}, \href
  {https://ui.adsabs.harvard.edu/abs/2019ApJ...887..173L} {887, 173}

\bibitem[\protect\citeauthoryear{{Lacy} et~al.,}{{Lacy}
  et~al.}{2004}]{Lacy2004ApJS}
{Lacy} M.,  et~al., 2004, \mn@doi [\apjs] {10.1086/422816}, \href
  {https://ui.adsabs.harvard.edu/abs/2004ApJS..154..166L} {154, 166}

\bibitem[\protect\citeauthoryear{{Lacy} et~al.,}{{Lacy}
  et~al.}{2013}]{Lacy2013ApJS}
{Lacy} M.,  et~al., 2013, \mn@doi [\apjs] {10.1088/0067-0049/208/2/24}, \href
  {https://ui.adsabs.harvard.edu/abs/2013ApJS..208...24L} {208, 24}

\bibitem[\protect\citeauthoryear{{Lanzuisi}, {Piconcelli}, {Fiore}, {Feruglio},
  {Vignali}, {Salvato}  \& {Gruppioni}}{{Lanzuisi} et~al.}{2009}]{Lanzuisi2009}
{Lanzuisi} G.,  {Piconcelli} E.,  {Fiore} F.,  {Feruglio} C.,  {Vignali} C.,
  {Salvato} M.,   {Gruppioni} C.,  2009, \mn@doi [\aap]
  {10.1051/0004-6361/200811282}, \href
  {https://ui.adsabs.harvard.edu/abs/2009A&A...498...67L} {498, 67}

\bibitem[\protect\citeauthoryear{{Lawrence}}{{Lawrence}}{1991}]{Lawrence1991MNRAS}
{Lawrence} A.,  1991, \mn@doi [\mnras] {10.1093/mnras/252.4.586}, \href
  {https://ui.adsabs.harvard.edu/abs/1991MNRAS.252..586L} {252, 586}

\bibitem[\protect\citeauthoryear{{Lawrence} \& {Elvis}}{{Lawrence} \&
  {Elvis}}{2010}]{LawrenceElvis2010}
{Lawrence} A.,  {Elvis} M.,  2010, \mn@doi [\apj]
  {10.1088/0004-637X/714/1/561}, \href
  {https://ui.adsabs.harvard.edu/abs/2010ApJ...714..561L} {714, 561}

\bibitem[\protect\citeauthoryear{{Liu}, {Zakamska}, {Greene}, {Nesvadba}  \&
  {Liu}}{{Liu} et~al.}{2013}]{Liu2013MNRAS436}
{Liu} G.,  {Zakamska} N.~L.,  {Greene} J.~E.,  {Nesvadba} N. P.~H.,   {Liu} X.,
   2013, \mn@doi [\mnras] {10.1093/mnras/stt1755}, \href
  {https://ui.adsabs.harvard.edu/abs/2013MNRAS.436.2576L} {436, 2576}

\bibitem[\protect\citeauthoryear{{Lonsdale} et~al.,}{{Lonsdale}
  et~al.}{2003}]{Lonsdale2003}
{Lonsdale} C.~J.,  et~al., 2003, \mn@doi [\pasp] {10.1086/376850}, \href
  {https://ui.adsabs.harvard.edu/abs/2003PASP..115..897L} {115, 897}

\bibitem[\protect\citeauthoryear{{Lusso} et~al.,}{{Lusso}
  et~al.}{2013}]{Lusso2013ApJ777}
{Lusso} E.,  et~al., 2013, \mn@doi [\apj] {10.1088/0004-637X/777/2/86}, \href
  {https://ui.adsabs.harvard.edu/abs/2013ApJ...777...86L} {777, 86}

\bibitem[\protect\citeauthoryear{{Mauduit} et~al.,}{{Mauduit}
  et~al.}{2012}]{Mauduit2012}
{Mauduit} J.~C.,  et~al., 2012, \mn@doi [\pasp] {10.1086/666945}, \href
  {https://ui.adsabs.harvard.edu/abs/2012PASP..124..714M} {124, 714}

\bibitem[\protect\citeauthoryear{{Osterbrock} \& {Ferland}}{{Osterbrock} \&
  {Ferland}}{2006}]{OsterbrockFerland2006}
{Osterbrock} D.~E.,  {Ferland} G.~J.,  2006, {Astrophysics of gaseous nebulae
  and active galactic nuclei}

\bibitem[\protect\citeauthoryear{{Papovich} et~al.,}{{Papovich}
  et~al.}{2016}]{Papovich2016ApJS224}
{Papovich} C.,  et~al., 2016, \mn@doi [\apjs] {10.3847/0067-0049/224/2/28},
  \href {https://ui.adsabs.harvard.edu/abs/2016ApJS..224...28P} {224, 28}

\bibitem[\protect\citeauthoryear{{Perrotta}, {Hamann}, {Zakamska},
  {Alexandroff}, {Rupke}  \& {Wylezalek}}{{Perrotta}
  et~al.}{2019}]{Perrotta2019MNRAS}
{Perrotta} S.,  {Hamann} F.,  {Zakamska} N.~L.,  {Alexandroff} R.~M.,  {Rupke}
  D.,   {Wylezalek} D.,  2019, \mn@doi [\mnras] {10.1093/mnras/stz1993}, \href
  {https://ui.adsabs.harvard.edu/abs/2019MNRAS.488.4126P} {488, 4126}

\bibitem[\protect\citeauthoryear{{Polletta} et~al.,}{{Polletta}
  et~al.}{2007}]{Polletta2007ApJ}
{Polletta} M.,  et~al., 2007, \mn@doi [\apj] {10.1086/518113}, \href
  {https://ui.adsabs.harvard.edu/abs/2007ApJ...663...81P} {663, 81}

\bibitem[\protect\citeauthoryear{{Polletta}, {Weedman}, {H{\"o}nig},
  {Lonsdale}, {Smith}  \& {Houck}}{{Polletta} et~al.}{2008}]{Polletta2008}
{Polletta} M.,  {Weedman} D.,  {H{\"o}nig} S.,  {Lonsdale} C.~J.,  {Smith}
  H.~E.,   {Houck} J.,  2008, \mn@doi [\apj] {10.1086/524343}, \href
  {https://ui.adsabs.harvard.edu/abs/2008ApJ...675..960P} {675, 960}

\bibitem[\protect\citeauthoryear{{Prochaska} et~al.,}{{Prochaska}
  et~al.}{2013}]{Prochaska2013ApJ776}
{Prochaska} J.~X.,  et~al., 2013, \mn@doi [\apj] {10.1088/0004-637X/776/2/136},
  \href {https://ui.adsabs.harvard.edu/abs/2013ApJ...776..136P} {776, 136}

\bibitem[\protect\citeauthoryear{{Prochaska} et~al.,}{{Prochaska}
  et~al.}{2020a}]{pypeit:zenodo}
{Prochaska} J.~X.,  et~al., 2020a, {pypeit/PypeIt: Release 1.0.0},
  \mn@doi{10.5281/zenodo.3743493}

\bibitem[\protect\citeauthoryear{{Prochaska}, {Hennawi}, {Westfall}, {Cooke},
  {Wang}, {Hsyu}, {Davies}  \& {Farina}}{{Prochaska}
  et~al.}{2020b}]{pypeit:joss_arXiv}
{Prochaska} J.~X.,  {Hennawi} J.~F.,  {Westfall} K.~B.,  {Cooke} R.~J.,  {Wang}
  F.,  {Hsyu} T.,  {Davies} F.~B.,   {Farina} E.~P.,  2020b, arXiv e-prints,
  \href {https://ui.adsabs.harvard.edu/abs/2020arXiv200506505P} {p.
  arXiv:2005.06505}

\bibitem[\protect\citeauthoryear{Prochaska et~al.,}{Prochaska
  et~al.}{2020c}]{pypeit:joss_pub}
Prochaska J.~X.,  et~al., 2020c, \mn@doi [Journal of Open Source Software]
  {10.21105/joss.02308}, 5, 2308

\bibitem[\protect\citeauthoryear{{Reyes} et~al.,}{{Reyes}
  et~al.}{2008}]{reyes2008AJ}
{Reyes} R.,  et~al., 2008, \mn@doi [\aj] {10.1088/0004-6256/136/6/2373}, \href
  {https://ui.adsabs.harvard.edu/abs/2008AJ....136.2373R} {136, 2373}

\bibitem[\protect\citeauthoryear{{Ricci} et~al.,}{{Ricci}
  et~al.}{2017}]{Ricci2017}
{Ricci} C.,  et~al., 2017, \mn@doi [\apj] {10.3847/1538-4357/835/1/105}, \href
  {https://ui.adsabs.harvard.edu/abs/2017ApJ...835..105R} {835, 105}

\bibitem[\protect\citeauthoryear{{Richards} et~al.,}{{Richards}
  et~al.}{2006}]{Richards2006}
{Richards} G.~T.,  et~al., 2006, \mn@doi [\aj] {10.1086/503559}, \href
  {https://ui.adsabs.harvard.edu/abs/2006AJ....131.2766R} {131, 2766}

\bibitem[\protect\citeauthoryear{{Riffel} et~al.,}{{Riffel}
  et~al.}{2021}]{Riffel2021}
{Riffel} R.,  et~al., 2021, \mn@doi [\mnras] {10.1093/mnras/staa3907}, \href
  {https://ui.adsabs.harvard.edu/abs/2021MNRAS.501.4064R} {501, 4064}

\bibitem[\protect\citeauthoryear{{Ross} et~al.,}{{Ross}
  et~al.}{2013}]{Ross2013}
{Ross} N.~P.,  et~al., 2013, \mn@doi [\apj] {10.1088/0004-637X/773/1/14}, \href
  {https://ui.adsabs.harvard.edu/abs/2013ApJ...773...14R} {773, 14}

\bibitem[\protect\citeauthoryear{{Ross} et~al.,}{{Ross}
  et~al.}{2015}]{Ross2015MNRAS453}
{Ross} N.~P.,  et~al., 2015, \mn@doi [\mnras] {10.1093/mnras/stv1710}, \href
  {https://ui.adsabs.harvard.edu/abs/2015MNRAS.453.3932R} {453, 3932}

\bibitem[\protect\citeauthoryear{{Sanders}, {Soifer}, {Elias}, {Madore},
  {Matthews}, {Neugebauer}  \& {Scoville}}{{Sanders}
  et~al.}{1988}]{Sanders1988ApJ}
{Sanders} D.~B.,  {Soifer} B.~T.,  {Elias} J.~H.,  {Madore} B.~F.,  {Matthews}
  K.,  {Neugebauer} G.,   {Scoville} N.~Z.,  1988, \mn@doi [\apj]
  {10.1086/165983}, \href
  {https://ui.adsabs.harvard.edu/abs/1988ApJ...325...74S} {325, 74}

\bibitem[\protect\citeauthoryear{{Sanders} et~al.,}{{Sanders}
  et~al.}{2007}]{Sanders2007}
{Sanders} D.~B.,  et~al., 2007, \mn@doi [\apjs] {10.1086/517885}, \href
  {https://ui.adsabs.harvard.edu/abs/2007ApJS..172...86S} {172, 86}

\bibitem[\protect\citeauthoryear{{Shen}}{{Shen}}{2016}]{Shen2016}
{Shen} Y.,  2016, \mn@doi [\apj] {10.3847/0004-637X/817/1/55}, \href
  {https://ui.adsabs.harvard.edu/abs/2016ApJ...817...55S} {817, 55}

\bibitem[\protect\citeauthoryear{{Soltan}}{{Soltan}}{1982}]{Soltan1982MNRAS}
{Soltan} A.,  1982, \mn@doi [\mnras] {10.1093/mnras/200.1.115}, \href
  {https://ui.adsabs.harvard.edu/abs/1982MNRAS.200..115S} {200, 115}

\bibitem[\protect\citeauthoryear{{Stern} et~al.,}{{Stern}
  et~al.}{2005}]{Stern2005ApJ}
{Stern} D.,  et~al., 2005, \mn@doi [\apj] {10.1086/432523}, \href
  {https://ui.adsabs.harvard.edu/abs/2005ApJ...631..163S} {631, 163}

\bibitem[\protect\citeauthoryear{{Temple}, {Banerji}, {Hewett}, {Coatman},
  {Maddox}  \& {Peroux}}{{Temple} et~al.}{2019}]{Temple2019MNRAS}
{Temple} M.~J.,  {Banerji} M.,  {Hewett} P.~C.,  {Coatman} L.,  {Maddox} N.,
  {Peroux} C.,  2019, \mn@doi [\mnras] {10.1093/mnras/stz1420}, \href
  {https://ui.adsabs.harvard.edu/abs/2019MNRAS.487.2594T} {487, 2594}

\bibitem[\protect\citeauthoryear{{Timlin} et~al.,}{{Timlin}
  et~al.}{2016}]{Timlin2016ApJS}
{Timlin} J.~D.,  et~al., 2016, \mn@doi [\apjs] {10.3847/0067-0049/225/1/1},
  \href {https://ui.adsabs.harvard.edu/abs/2016ApJS..225....1T} {225, 1}

\bibitem[\protect\citeauthoryear{{Treister} \& {Urry}}{{Treister} \&
  {Urry}}{2005}]{Treister2005}
{Treister} E.,  {Urry} C.~M.,  2005, \mn@doi [\apj] {10.1086/431892}, \href
  {https://ui.adsabs.harvard.edu/abs/2005ApJ...630..115T} {630, 115}

\bibitem[\protect\citeauthoryear{{Tsai} et~al.,}{{Tsai}
  et~al.}{2015}]{Tsai2015}
{Tsai} C.-W.,  et~al., 2015, \mn@doi [\apj] {10.1088/0004-637X/805/2/90}, \href
  {https://ui.adsabs.harvard.edu/abs/2015ApJ...805...90T} {805, 90}

\bibitem[\protect\citeauthoryear{{Ueda}, {Akiyama}, {Ohta}  \& {Miyaji}}{{Ueda}
  et~al.}{2003}]{ueda2003}
{Ueda} Y.,  {Akiyama} M.,  {Ohta} K.,   {Miyaji} T.,  2003, \mn@doi [\apj]
  {10.1086/378940}, \href
  {https://ui.adsabs.harvard.edu/abs/2003ApJ...598..886U} {598, 886}

\bibitem[\protect\citeauthoryear{{Vestergaard} \& {Peterson}}{{Vestergaard} \&
  {Peterson}}{2006}]{Vestergaard2006}
{Vestergaard} M.,  {Peterson} B.~M.,  2006, \mn@doi [\apj] {10.1086/500572},
  \href {https://ui.adsabs.harvard.edu/abs/2006ApJ...641..689V} {641, 689}

\bibitem[\protect\citeauthoryear{{Vito} et~al.,}{{Vito}
  et~al.}{2018}]{Vito2018}
{Vito} F.,  et~al., 2018, \mn@doi [\mnras] {10.1093/mnras/stx3120}, \href
  {https://ui.adsabs.harvard.edu/abs/2018MNRAS.474.4528V} {474, 4528}

\bibitem[\protect\citeauthoryear{{Werner} et~al.,}{{Werner}
  et~al.}{2004}]{Werner2004}
{Werner} M.~W.,  et~al., 2004, \mn@doi [\apjs] {10.1086/422992}, \href
  {https://ui.adsabs.harvard.edu/abs/2004ApJS..154....1W} {154, 1}

\bibitem[\protect\citeauthoryear{{Wright} et~al.,}{{Wright}
  et~al.}{2010}]{Wright2010}
{Wright} E.~L.,  et~al., 2010, \mn@doi [\aj] {10.1088/0004-6256/140/6/1868},
  \href {https://ui.adsabs.harvard.edu/abs/2010AJ....140.1868W} {140, 1868}

\bibitem[\protect\citeauthoryear{{Wu} et~al.,}{{Wu} et~al.}{2012}]{Wu2012ApJ}
{Wu} J.,  et~al., 2012, \mn@doi [\apj] {10.1088/0004-637X/756/1/96}, \href
  {https://ui.adsabs.harvard.edu/abs/2012ApJ...756...96W} {756, 96}

\bibitem[\protect\citeauthoryear{{Wu} et~al.,}{{Wu} et~al.}{2018}]{Wu2018}
{Wu} J.,  et~al., 2018, \mn@doi [\apj] {10.3847/1538-4357/aa9ff3}, \href
  {https://ui.adsabs.harvard.edu/abs/2018ApJ...852...96W} {852, 96}

\bibitem[\protect\citeauthoryear{{Yan} et~al.,}{{Yan}
  et~al.}{2013}]{Yan2013AJ145}
{Yan} L.,  et~al., 2013, \mn@doi [\aj] {10.1088/0004-6256/145/3/55}, \href
  {https://ui.adsabs.harvard.edu/abs/2013AJ....145...55Y} {145, 55}

\bibitem[\protect\citeauthoryear{{York} et~al.,}{{York}
  et~al.}{2000}]{york2000}
{York} D.~G.,  et~al., 2000, \mn@doi [\aj] {10.1086/301513}, \href
  {https://ui.adsabs.harvard.edu/abs/2000AJ....120.1579Y} {120, 1579}

\bibitem[\protect\citeauthoryear{{Yu} \& {Tremaine}}{{Yu} \&
  {Tremaine}}{2002}]{YuTremaine2002}
{Yu} Q.,  {Tremaine} S.,  2002, \mn@doi [\mnras]
  {10.1046/j.1365-8711.2002.05532.x}, \href
  {https://ui.adsabs.harvard.edu/abs/2002MNRAS.335..965Y} {335, 965}

\bibitem[\protect\citeauthoryear{{Yuan}, {Strauss}  \& {Zakamska}}{{Yuan}
  et~al.}{2016}]{Yuan2016MNRAS}
{Yuan} S.,  {Strauss} M.~A.,   {Zakamska} N.~L.,  2016, \mn@doi [\mnras]
  {10.1093/mnras/stw1747}, \href
  {https://ui.adsabs.harvard.edu/abs/2016MNRAS.462.1603Y} {462, 1603}

\bibitem[\protect\citeauthoryear{{Zakamska} \& {Greene}}{{Zakamska} \&
  {Greene}}{2014}]{ZakamskaGreene2014}
{Zakamska} N.~L.,  {Greene} J.~E.,  2014, \mn@doi [\mnras]
  {10.1093/mnras/stu842}, \href
  {https://ui.adsabs.harvard.edu/abs/2014MNRAS.442..784Z} {442, 784}

\bibitem[\protect\citeauthoryear{{Zakamska} et~al.,}{{Zakamska}
  et~al.}{2003}]{zakamska2003AJ}
{Zakamska} N.~L.,  et~al., 2003, \mn@doi [\aj] {10.1086/378610}, \href
  {https://ui.adsabs.harvard.edu/abs/2003AJ....126.2125Z} {126, 2125}

\bibitem[\protect\citeauthoryear{{Zakamska} et~al.,}{{Zakamska}
  et~al.}{2006}]{Zakamska2006}
{Zakamska} N.~L.,  et~al., 2006, \mn@doi [\aj] {10.1086/506986}, \href
  {https://ui.adsabs.harvard.edu/abs/2006AJ....132.1496Z} {132, 1496}

\bibitem[\protect\citeauthoryear{{Zakamska} et~al.,}{{Zakamska}
  et~al.}{2016}]{Zakamska2016b}
{Zakamska} N.~L.,  et~al., 2016, \mn@doi [\mnras] {10.1093/mnras/stw718}, \href
  {https://ui.adsabs.harvard.edu/abs/2016MNRAS.459.3144Z} {459, 3144}

\bibitem[\protect\citeauthoryear{{Zakamska} et~al.,}{{Zakamska}
  et~al.}{2019}]{Zakamska2019}
{Zakamska} N.~L.,  et~al., 2019, \mn@doi [\mnras] {10.1093/mnras/stz2071},
  \href {https://ui.adsabs.harvard.edu/abs/2019MNRAS.489..497Z} {489, 497}

\bibitem[\protect\citeauthoryear{{Zappacosta} et~al.,}{{Zappacosta}
  et~al.}{2018}]{Zappacosta2018}
{Zappacosta} L.,  et~al., 2018, \mn@doi [\aap] {10.1051/0004-6361/201732557},
  \href {https://ui.adsabs.harvard.edu/abs/2018A&A...618A..28Z} {618, A28}

\makeatother
\end{thebibliography}







\bsp	
\label{lastpage}
\end{document}